\documentclass{article}[10pt]
\usepackage[utf8]{inputenc}

\usepackage{pgfplots}
\pgfplotsset{compat=1.9}
\usepackage{tikz}
\usetikzlibrary{arrows}
\usepackage{longtable}
\usepackage{threeparttable}
\usepackage{caption}
\usepackage{pdflscape}
\usepackage{amsmath}
\usepackage[super]{nth}
\usepackage{float}
\usepackage{verbatim}
\usepackage{adjustbox}
\usepackage{rotate}
\usepackage{amssymb}
\usepackage{amsmath} \usepackage{amsthm} \usepackage{a4} \usepackage{marvosym} \usepackage{parskip}
\usepackage{graphicx}
\usepackage{rotating} 
\usepackage{natbib}
\usepackage{pstricks, pst-all, pst-tree} 
\usepackage{eurosym}
\usepackage{threeparttablex}
\usepackage[a4paper, left=2.5cm, right=2.5cm, top=2.5cm, bottom=2.5cm]{geometry}
\usepackage{graphicx}

\usepackage{tikz}
\usepackage{comment}
\usepackage{booktabs}
\usepackage{setspace}
\usepackage[pdfauthor={},pdftitle={}, pdfstartview={FitH}]{hyperref}
\usepackage{subcaption}
\usepackage{array} 
\newcolumntype{C}[1]{>{\centering\arraybackslash}p{#1}}

\usepackage{soul}

\theoremstyle{definition}

\theoremstyle{definition}
\theoremstyle{definition}
\theoremstyle{definition}

\usepackage{array}
\newcolumntype{H}{>{\setbox0=\hbox\bgroup}c<{\egroup}@{}}
\begin{document}

\title{\Large Pollution and Mortality: Evidence from early \nth{20}-Century Sweden\thanks{Haylock: CINCH, University of Duisburg-Essen, Weststadtt\"{u}rme Berliner Platz 6-8, 45127 Essen, Germany; e-mail: michael.haylock@uni-due.de;Karlsson: CINCH, University of Duisburg-Essen, Weststadtt\"{u}rme Berliner Platz 6-8, 45127 Essen, Germany; e-mail: martin.karlsson@uni-due.de; Obrizan: Economics Department, Kyiv School of Economics, Shpaka 3 Str, 39600, Kyiv, Ukraine; e-mail: mobrizan@kse.org.ua. A research grant from the German Research Foundation DFG (Project number 467132381) is gratefully acknowledged.}}
\author{Michael Haylock  \;\; Martin Karlsson  \;\; Maksym Obrizan}

\maketitle
\thispagestyle{empty}

\onehalfspacing
\begin{abstract}
Economic growth in Sweden during the early 20th Century was largely driven by industry. A significant contributor to this growth was the installation of different kinds of engines used to power factories. We use newly digitized data on engines and their energy source by industry sector, and combine this with municipality-level data of workers per industry sector to construct a new variable reflecting economic output using dirty engines. In turn, we assess the average externality of dirty output on mortality in the short-run, as defined by deaths over the population in the baseline year. Our results show substantial increases of up to 17\% higher mortality in cities where large increases to dirty engine installations occurred, which is largely driven by the elderly. We also run a placebo test using clean powered industry and find no effect on mortality.
\end{abstract}

\newpage
\onehalfspacing
\setcounter{page}{1} 

\section{Introduction}

Over the course of the \nth{20} century, life expectancy experienced a dramatic increase, representing one of the most remarkable achievements in human history. This improvement has been attributed to a combination of factors, including public health investments such as hospitals and sanitation systems \citep{hollingsworth24,alsan2019watersheds,anderson2022reexamining}, disease eradication campaigns targeting tuberculosis and other infectious diseases \citep{anderson2019first,egedeso2020preventing}, and advances in medical technology, such as antibiotics and vaccines \citep{cutler2006determinants,jayachandran2010modern,ager2023medical}. Improvements in nutrition and rising incomes also played a significant role in reducing mortality and improving overall health outcomes \citep{costa2015health}.

Amid these advances, environmental factors such as pollution have emerged as an important determinant of health and mortality. The rapid industrialisation of the late \nth{19} and early \nth{20} centuries brought both economic progress and significant environmental costs, with air pollution from coal and other fossil fuels playing a particularly harmful role. Historical studies, while less abundant than modern analyses, shed light on the mortality consequences of pollution in this transformative period. Understanding the interplay between industrial growth, environmental degradation, and public health is crucial for a more comprehensive account of the factors driving improvements in life expectancy. Our paper contributes to this literature by focusing on the short-run impact of pollution on mortality during a key period of industrialization in Sweden. By doing so, we aim to illuminate a neglected dimension of the historical determinants of health and to provide insights relevant for understanding contemporary challenges in developing economies.


We assess the impact of pollution, driven by increased industrial demand, on short-term mortality in urban areas during the early \nth{20} century. To do so, we develop a novel proxy for pollution, identifying its effects through sudden increases in the number of workers in pollution-intensive industries at the city level. While economic growth is associated with rising energy consumption, which could independently affect health and mortality, these effects are likely to work in the opposite direction by improving incomes, access to medical care, and overall living conditions. As a result, economic growth could create countervailing effects that lead to an underestimation of the true impact of pollution on mortality. Importantly, the mechanisms linked to economic growth tend to operate over the medium to long term. To address these endogeneity concerns, we exploit the fact that several Swedish towns had abundant access to hydropower—a clean and renewable energy source. By leveraging this natural variation, we can effectively disentangle the impact of pollution from the broader effects of economic growth, providing a credible estimate of pollution's effect on mortality.

The Industrial Revolution brought remarkable economic progress but also significant environmental costs, particularly from air pollution caused by the adoption of steam engines and coal combustion. In \nth{19}-century London, pollution levels exceeded 600 micrograms per cubic meter, surpassing even modern-day Delhi's notorious levels of under 400 micrograms per cubic meter \citep{brimblecombe1987big, fouquet2011long}. Such high pollution levels contributed to severe health consequences, including respiratory illnesses and elevated mortality rates, as noted in historical accounts and early research on air pollution \citep{brimblecombe1998history}.

The link between air pollution and mortality has been a persistent theme in research. Early studies often focused on extreme pollution events, such as London’s 1952 ``killer fog,'' which coincided with a significant rise in mortality \citep{logan1953mortality}. Subsequent studies in modern settings, such as the Harvard Six Cities Study \citep{dockery1993association}, revealed robust associations between long-term exposure to air pollution and mortality, findings confirmed by reanalyses \citep{krewski2005reanalysis}. These studies highlight the persistent threat of air pollution in advanced economies.

Historical studies remain highly relevant for understanding air pollution's effects, as many developing economies today experience pollution levels similar to those of advanced economies during industrialisation. Research in historical contexts provides key insights into the health impacts of pollution at levels rarely seen in modern high-income countries. For instance, \citet{clay2010did} documented how regulatory and technological changes in late-\nth{19}-century London reduced coal smoke emissions, leading to fewer foggy days and declines in fog-related mortality. Similarly, \citet{hanlon2015pollution} quantified the effects of industrial pollution in \nth{19}-century Britain, showing that increases in coal-based pollution raised mortality rates by amounts comparable to major infectious diseases like smallpox. Other studies have linked coal consumption to tuberculosis in various historical contexts, including Canada, the U.S., China, Norway, and Japan \citep{tremblay2007historical}, and highlighted the exacerbating role of air pollution during the 1918 Spanish Influenza Pandemic \citep{clay2018pollution}.

Despite these contributions, the historical literature remains limited in scope and data quality compared to modern studies. Our paper addresses these gaps and advances the literature in several ways. First, we use individual-level mortality data for entire urban populations, enabling us to derive credible estimates not only for the urban population as a whole but also for specific subgroups. Second, leveraging detailed administrative data on local energy production, we estimate a clear dose-response relationship that strengthens identification. Third, this data allows us to construct a pollution proxy that distinguishes sudden jumps in pollution-generating activities from secular trends. Fourth, we exploit excellent data on hydropower to address potential confounding from general economic growth, providing credible estimates of pollution's effects on mortality.

By synthesising these contributions, our study bridges the gap between historical and modern research on air pollution and health. The paper is structured as follows: Section \ref{section2} provides the historical context, Section \ref{section3} outlines our empirical strategy and data, and Section \ref{section4} presents our main results and robustness checks.

\section{Historical Context}\label{section2}

The first half of the \nth{20} century marked a transformative period for Sweden, as the country transitioned from an agrarian society to an industrialized economy. By 1900, 53\% of the Swedish population still derived their livelihood from agriculture, but industrialization was rapidly gaining momentum. By 1930, the share of the population employed in agriculture had declined to 39.4\%, while manufacturing had risen to 35.7\% \citep{sweden1942folkrakningen}. This shift reflected the broader economic transformation, as industrial output overtook agricultural production in absolute terms by 1910 \citep{schon2010sweden}. Between 1890 and 1910, industrial growth averaged 5.5\% annually, fueled by both domestic demand for consumer goods and expanding trade with major export markets like Britain, Germany, and Scandinavian neighbours. Urban centers such as Stockholm, Gothenburg, and Malmö emerged as hubs of industrial activity, shaping Sweden's economic landscape.

During World War I, Sweden’s economy faced disruptions due to its neutrality stance and the British-imposed naval blockade of the North Sea, which limited imports from overseas \citep{jorberg1978ekonomisk}. Domestically, this led to increased regulation, including price controls on essential goods, food rationing, and restricted access to certain materials \citep{schon2010sweden}. Nonetheless, the war had mixed effects on the economy: it spurred a surge in exports as Sweden stepped into the void left by foreign competitors. Agriculture also benefited, and the overall economic environment was favourable for Swedish industries. This period of economic growth was supported by a highly liquid capital market, fueled by increased long-term savings and the establishment of new insurance companies and local banks (Larsson, 1998). However, the war resulted in significant redistribution of wealth, favouring capital owners over workers \citep{schon2010sweden}.

The post-war boom was interrupted by the economic downturn of 1920–21, during which GDP declined by 5\% and unemployment surged. While recovery was relatively swift, the sectors that had thrived during the war were the hardest hit. The 1920s then saw fast growth in real wages and a shift toward higher returns to labor relative to capital, partly due to the introduction of shorter working hours \citep{magnusson2010sveriges,schon2010sweden}.

The rapid industrial expansion of the early \nth{20} century brought both economic gains and environmental challenges, especially concerning pollution in urban areas. Industrial production, particularly in sectors such as iron, steel, textiles, and timber, relied heavily on coal and other fossil fuels, leading to high levels of air pollution. Factories emitted significant quantities of particulate matter and sulphur dioxide, which, combined with emissions from residential heating and transportation, degraded urban air quality. In densely populated cities, these pollutants posed serious health risks, with respiratory ailments becoming more common among urban residents and factory workers.

In response to the need for consistent fuel sources, industries in Sweden used a range of materials, including coal, wood waste, charcoal, and even alum shale, although consumption patterns varied widely between sectors. For instance, the paper and graphics industry accounted for approximately 34.5\% of the recorded industrial fuel consumption, while the mining and metal industries consumed about 19.3\% \citep{1920i}. Despite efforts to document fuel use comprehensively, smaller enterprises, particularly those relying on production by-products, often provided incomplete data. In total, Sweden’s industrial fuel consumption reached an estimated 3,305,041 tons of coal equivalent in the early 1920s, underscoring the significant environmental footprint of this period's rapid industrial growth \citep{1922i}.

\begin{figure}[H]
    \centering
    \begin{subfigure}[b]{0.45\linewidth}
        \centering
        \includegraphics[width=\linewidth]{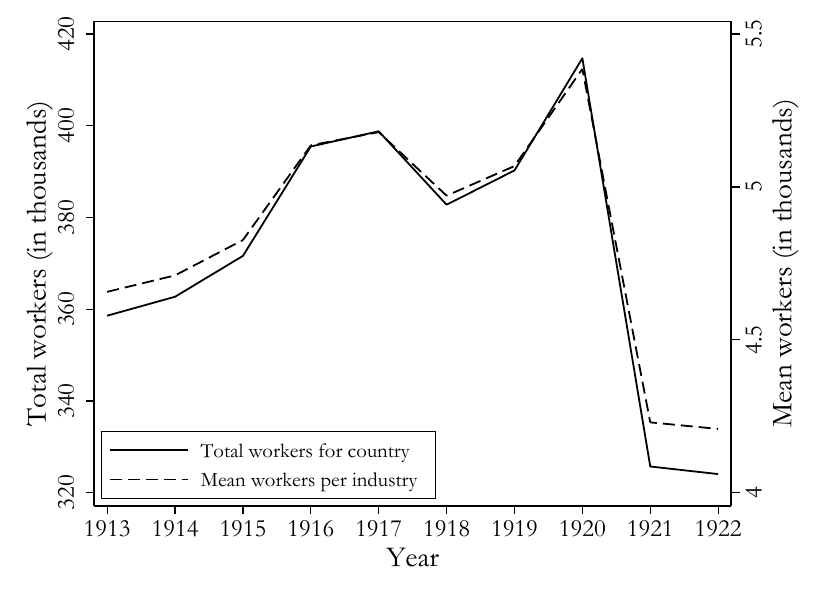}
        \caption{City-Level Industrial Workforce}\label{fig:workforce}
    \end{subfigure}
    \begin{subfigure}[b]{0.45\linewidth}
        \centering
        \includegraphics[width=\linewidth]{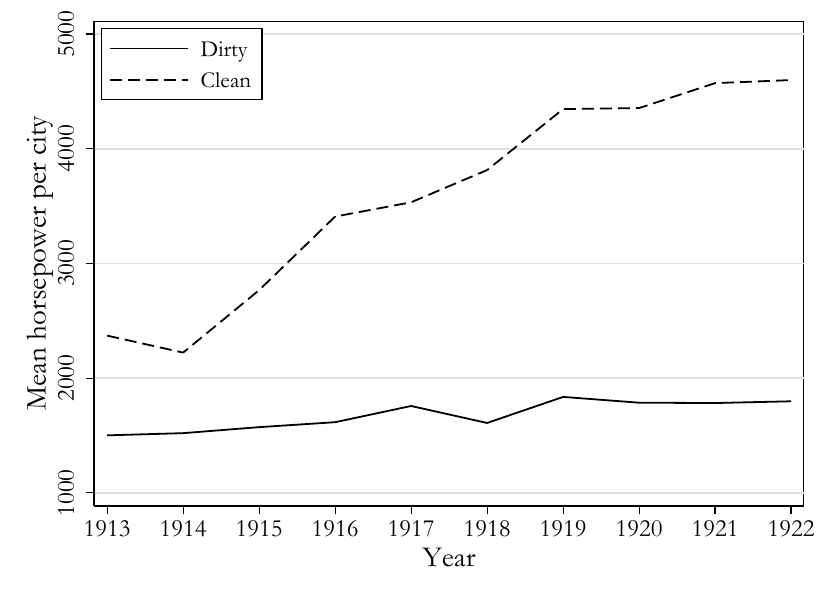}
        \caption{Energy Generation}\label{fig:energy}
    \end{subfigure}
    \caption{Trends in Industrial Workforce and Energy Sources, 1913–1921.}
    \label{fig:worker_desc}
       \caption*{\scriptsize \textit{Notes:} Own calculations based on data from the annual factory census \citep{1913i}.}
\end{figure}

Figure \ref{fig:worker_desc} provides descriptives concerning the steady trend of industrial growth and the dramatic economic fluctuations of the early 1920's. Figure \ref{fig:workforce} illustrates the total number of workers in the workforce at the national level and the average number of workers per industry. This reflects a steady increase in industrial employment over the sample period, resulting in a 14\% rise from 1913 to 1920, with over 5,000 additional workers entering the sector. However, the economic downturn of 1920–21 disrupted this trend, emphasising the volatility of the period.

Figure \ref{fig:energy} highlights the growing adoption of clean energy, such as water wheels and water turbines, across cities. On average, these sources exhibited much larger increases than ``dirty'' energy sources like coal-powered engines. This shift toward hydropower suggests that clean energy played a critical role in sustaining economic growth during this transformative period. The observed patterns also provide a foundation for placebo tests, allowing us to explore whether short-term economic growth impacts mortality.


\section{Empirical Strategy and Data}\label{section3}

\subsection{Identification Strategy}

Analysing the impact of pollution on health outcomes is challenging, given the endogenous nature of pollution. Pollution exposure is not random but likely correlates with a number of potentially confounding variables at the individual level, such as e.g. socio-economic status or pre-existing health conditions. Moreover, in this period, exposure to pollution often mirrored economic growth: the rapid industrialisation that Sweden was undergoing brought both higher incomes and increased pollution. As a result, disentangling the effects of pollution from the benefits of economic progress becomes complex, as pollution may reflect broader trends in Sweden's economic development.

Our main strategy to address concerns about endogeneity and confounding factors is to leverage sudden changes in pollution exposure driven by the local expansion of heavily polluting industries. We construct a pollution indicator at the city-year level for a sample of 95 Swedish towns and cities, based on detailed data on industrial capacity from an annual factory census \citep{1913i}.

To begin, we describe the sample and the source and construction of key variables. Our main sample comprises all towns and cities of Sweden, covered in the factory census yearbook from 1913 to 1922 \citep{1913i}. From this yearbook, we extract the number of workers in each city by industry sector, defining our variables based on the second-digit level. Industries are divided into nine broad categories, including e.g. `Mining and Metal Industry', `Wood Industry', and `Chemical and Technical Industry'. Each of these major categories is further subdivided into around ten second-digit categories, such as `Iron Ore Mines and Enrichment Plants', `Iron and Steel Manufacturing Plants', and `Mechanical Workshops'. The classification remains consistent across all years. Additionally, the factory census provides data on the power sources used in each industry at the national level, measured in horsepower. Power types include water wheels, water turbines, steam engines, steam turbines, oil engines, and gas engines.

To estimate the pollution potential, we calculate a ``dirty horsepower''  measure for each industry across the country by summing all horsepower types except for water wheels and water turbines. Using this measure, we derive a pollution proxy for each city as follows: First, we calculate the average dirty horsepower per worker at the national level for each second-digit industry. Next, for each city, we multiply the number of workers in a specific industry and year by the average dirty horsepower for that industry. Finally, we sum the dirty horsepower across all industries for each city and year, creating a proxy for pollution intensity in each city-year observation.

Our final dataset includes 1,027 city-year observations, forming an unbalanced panel due to some cities being newly founded during our study period. In robustness checks, we also examine a balanced panel of cities. Rural areas are also covered in the census but since the location of factories is not provided, they are excluded from the analysis. Figure \ref{fig:individual_dirty} shows the distribution of logs for individual energy sources, while Figure \ref{fig:sumdirty_dist} displays the distribution of our pollution proxy, including both the total dirty horsepower and its logarithmic transformation.

To gain a basic understanding of the functional-form relationship between pollution exposure and mortality, we begin by estimating the coefficients associated with the deciles of our dirty horsepower measure. An overview of the deciles and their associated amount of ``dirty horsepower'' is provided in Table \ref{tab:summary_sum_dirty}.

\begin{table}[H]
    \centering
    \caption{Summary of dirty energy capacity by decile}
    \label{tab:summary_sum_dirty}
    \begin{tabular}{cccc}
        \hline
        Decile & Mean & Std. Dev. & Freq. \\ 
        \hline
        1 & 27.886774 & 20.139928 & 103 \\ 
        2 & 110.57552 & 30.253093 & 103 \\ 
        3 & 230.6651 & 39.364348 & 103 \\ 
        4 & 389.89737 & 56.270154 & 102 \\ 
        5 & 579.87817 & 63.45166 & 103 \\ 
        6 & 845.69553 & 102.02904 & 103 \\ 
        7 & 1180.7556 & 83.443816 & 102 \\ 
        8 & 1506.0998 & 122.69018 & 103 \\ 
        9 & 2183.0532 & 302.68336 & 103 \\ 
        10 & 9834.9431 & 9617.7391 & 102 \\ 
        \hline
        Total & 1682.7729 & 4107.6502 & 1,027 \\ 
        \hline
    \end{tabular}
\end{table}

\begin{figure}[H]
       \centering
        \includegraphics[width=0.7\linewidth]{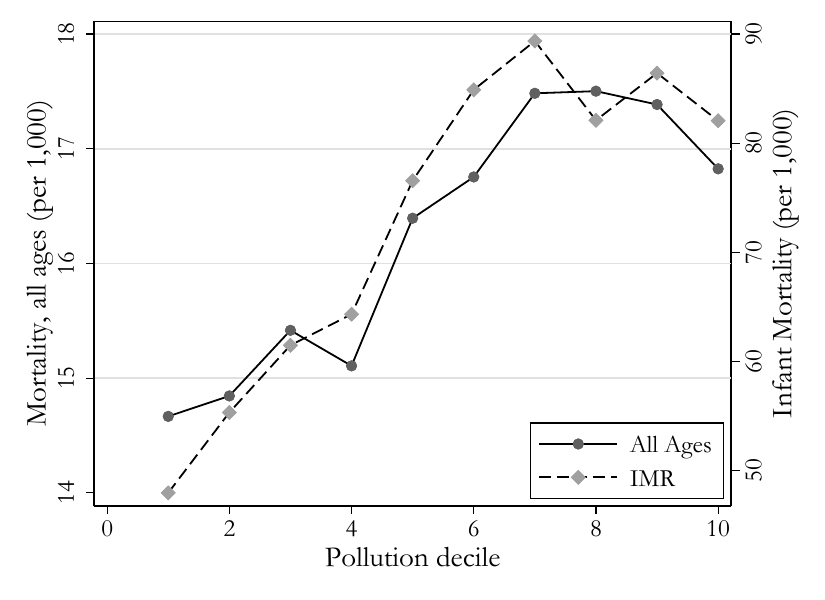}
\caption{Relationship between Mortality Rates and Pollution Exposure Deciles}
        \label{fig:MortGraph}
         \end{figure}

Using a simple two-way fixed-effects design, we regress city mortality rates on the previous year's pollution decile. The results, shown in Figure \ref{fig:MortGraph}, suggest a possible dose-response relationship for both overall and infant mortality. Specifically, the curves indicate an approximately linear relationship between deciles 1–5, after which the relationship flattens.\footnote{The full set of regression results is presented in Table \ref{tab:TWFE}.} In Figure \ref{fig:MortGraph2}, we further show that this implies a strongly non-linear relationship between mortality and actual pollution levels.

Based on these descriptives, we posit that most of the pollution-related variation in mortality will be captured by movements between the lower and the upper deciles. Our main treatment indicator is thus that a city reaches a level of DH in decile 5 or above for the first time.

\subsection{Outcome Data}

Our second key data source is the Swedish Deathbook, a comprehensive registry that records the birth and death dates of the entire Swedish population, covering the period from 1880 to the present \citep{fsgs_swedish_death_index_2022}. For this study, we focus on deaths that occurred between 1913 and 1922, aligning with the sample period of our analysis. We calculate the total number of deaths for each city-year cell, where the city is defined by the place of death. Additionally, we break down the deaths into age brackets: under 1 year old, 1 to 13 years, 14 to 29 years, 30 to 49 years, 50 to 74 years, and 75 years and older.

To calculate mortality rates, we divide the number of deaths in each age bracket by the population at baseline. The population data at baseline is drawn from a panel dataset of Swedish parishes, which provides detailed demographic information for the relevant time periods. In all analyses, mortality rates are expressed per 1,000 individuals to standardize the data and facilitate comparisons across age groups and time periods. This method allows us to accurately assess the mortality outcomes for different age groups within Swedish cities during the specified years.

Summary statistics for our main analysis sample (DiD sample) are provided in Table \ref{sum_did}. The full sample is summarized in Table \ref{sum_full}. The corresponding descriptives for the sample used in TWFE regressions are provided in Appendix Table \ref{sum_twfe}.

{
\def\sym#1{\ifmmode^{#1}\else\(^{#1}\)\fi}
\begin{longtable}{l*{1}{cccccc}}
\caption{Summary Statistics for DID sample}\\
\toprule\endfirsthead\midrule\endhead\midrule\endfoot\endlastfoot
                    &\multicolumn{6}{c}{}                                                         \\
                    &        Obs.&        Mean&        Med.&        S.D.&        Min.&        Max.\\
\midrule
Mortality all ages  &         460&      18.135&      15.291&       10.67&         4.4&        83.0\\
Mortality under 1   &         460&       1.455&       1.122&        1.33&         0&         9.6\\
Mortality 1 to 13   &         460&       1.440&       1.063&        1.35&         0&         8.4\\
Mortality 14 to 29  &         460&       2.010&       1.596&        1.65&         0&        10.5\\
Mortality 30 to 49  &         460&       2.274&       1.915&        1.63&         0&        12.4\\
Mortality 50 to 74  &         460&       5.882&       5.029&        3.69&         0.6&        27.2\\
Mortality 75 and over&         460&       5.074&       4.280&        3.73&         0&        25.4\\
Deaths all ages     &         460&      55.676&      43&       40.27&         7&       235\\
Deaths under 1      &         460&       4.904&       3&        5.26&         0&        33\\
Deaths 1 to 13      &         460&       4.722&       3&        5.33&         0&        34\\
Deaths 14 to 29     &         460&       6.459&       5&        5.96&         0&        40\\
Deaths 30 to 49     &         460&       7.296&       6&        6.18&         0&        35\\
Deaths 50 to 74     &         460&      17.800&      14&       13.08&         1&        77\\
Deaths 75 and over  &         460&      14.496&      12&       11.36&         0&        72\\
Population    &         460&   3,508.507&   2,919.500&    2,242.68&       839&     12972\\
Dirty HP            &         459&     258.947&     218.978&      211.24&         1.7&      1142.2\\
log dirty HP        &         459&       5.035&       5.389&        1.26&         0.5&         7.0\\
\bottomrule
\end{longtable}
\label{sum_did}
}

\subsection{Estimation Procedure}
To estimate the average treatment effects on the treated (ATTs), we use the difference-in-differences (DiD) methodology proposed by \citet{callaway2021difference}. We leverage never-treated cities as the control group, with no additional control variables included in the model in our main specification. By comparing mortality outcomes across cities that were treated at different times, we account for the temporal variation in exposure to the treatment, namely, the shock to factory capacity.

A key assumption in our analysis is that, in the absence of the shock to factory capacity, the treated cities would have followed a similar trend in mortality rates as the cities that were never treated or were treated at a later time. To support the plausibility of this assumption, we examine pre-treatment trends in mortality rates, comparing treated cities to cities that were not yet treated. These pre-treatment trends provide suggestive evidence that, had it not been for the shock, the mortality rates in treated cities would have evolved similarly to those in the control cities.

Additionally, we must assume that no other third factors, correlated with the treatment, influence both the shocks in factory capacity and mortality rates. This would ensure that the observed effects are not driven by confounding factors that could spuriously drive the treatment effect. To address concerns about work-related accidents, which could potentially bias our results, we assess the impacts separately for different age groups. This allows us to verify whether the treatment effect is specific to certain populations and is not simply due to external factors affecting mortality outcomes. Furthermore, we recognise the potential endogeneity of pollution to economic growth, which could confound the relationship between pollution and mortality. As discussed in the introduction, we address this concern by exploiting natural variation in access to hydropower in certain Swedish towns. This variation in clean energy use allows us to disentangle the impact of pollution from broader economic growth effects, providing a more accurate estimate of pollution’s impact on mortality.

\section{Results}\label{section4}

\subsection{First Stage}

In Figure \ref{fig:fstage}, we display the first-stage relationship between our treatment indicator and the amount of dirty horsepower (DH). The pollution shock corresponds to an increase in DH by 212 units in the year of the jump, or approximately 63 percent relative to the pretreatment mean of the treated group prior to the jump. Additionally, the event study shows that cities follow a parallel trend for at least four years prior to the treatment. The pollution shock itself persists for about four years before returning to the baseline.

\begin{figure}
       \centering
       \includegraphics[width=0.7\linewidth]{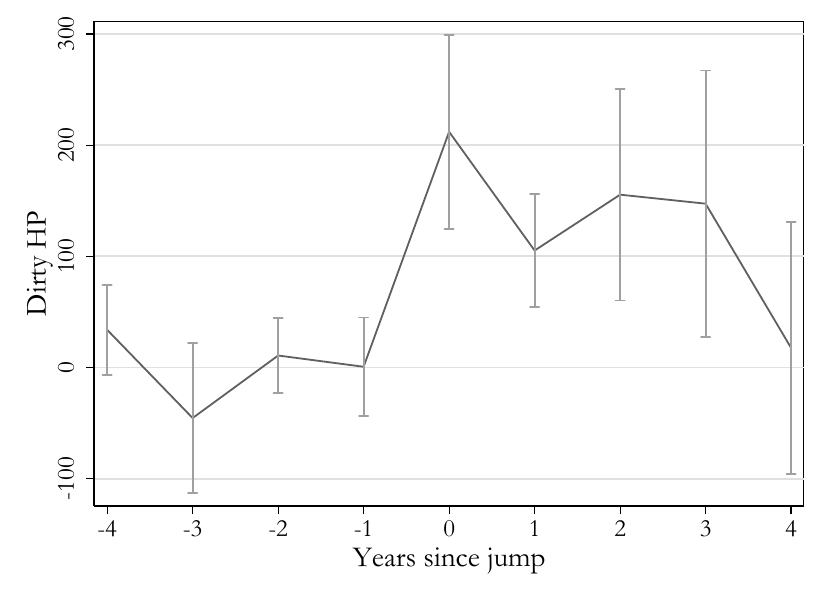}
\caption{First stage: Impact on Dirty Horsepowers}
        \label{fig:fstage}
         \end{figure}

\subsection{Effects on Mortality}

In Figure \ref{fig:mainmort}, we present the main result: the effects of the pollution shock on local mortality rates. The results indicate a gradual increase in mortality rates over the duration of the pollution shock, becoming statistically significant in the fourth year, with an estimated effect size of about 3 additional deaths per 1,000, or 16.8 percent relative to baseline mean of the treated group prior to treatment (17.8 deaths per 1,000). The overall ATT is 1.41 more deaths per 1,000 population. This corresponds to a substantial increase of approximately 8 percent over the baseline value.

\begin{figure}[H]
       \centering
       \includegraphics[width=0.7\linewidth]{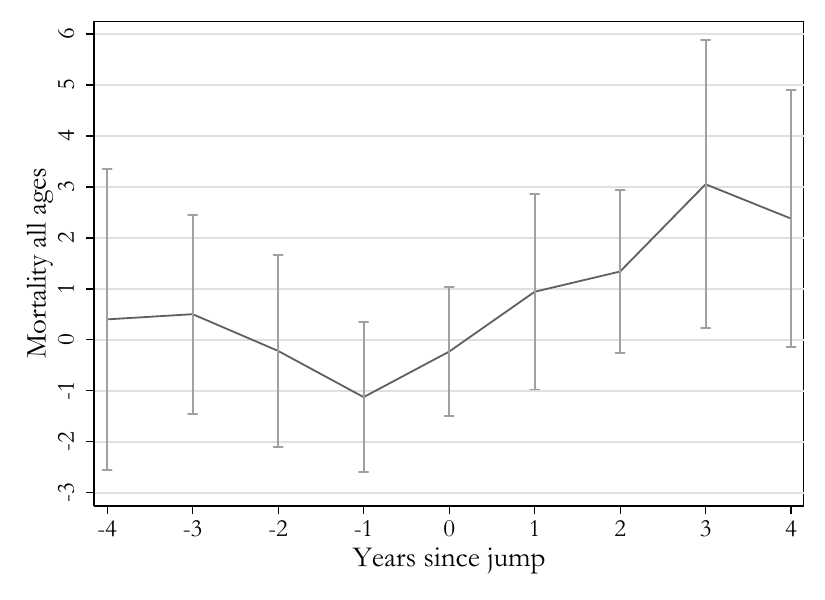}
\caption{Main Results: All-Age Mortality}
        \label{fig:mainmort}
         \end{figure}

\begin{figure}[H]
    \centering
    \begin{subfigure}[b]{0.45\linewidth}
        \centering
        \includegraphics[width=\linewidth]{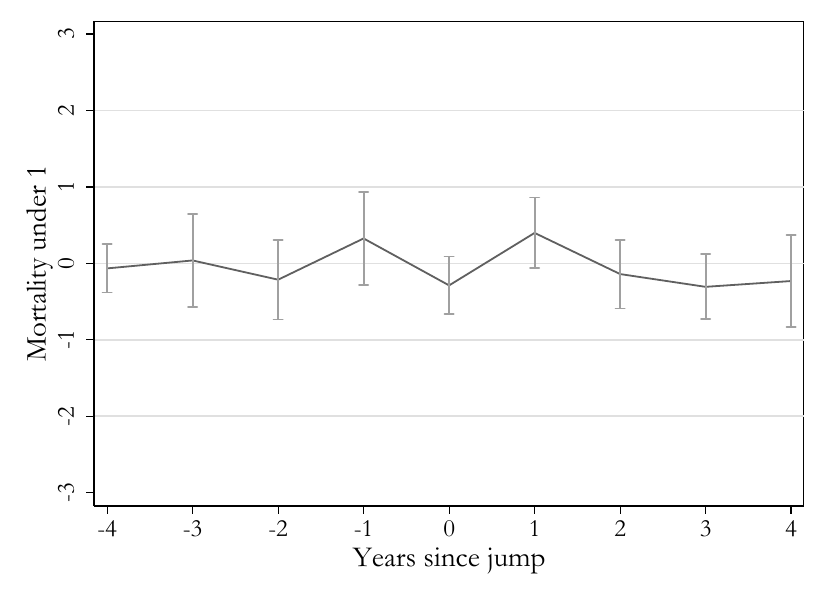}
        \caption{Infant Mortality}
    \end{subfigure}
    \begin{subfigure}[b]{0.45\linewidth}
        \centering
        \includegraphics[width=\linewidth]{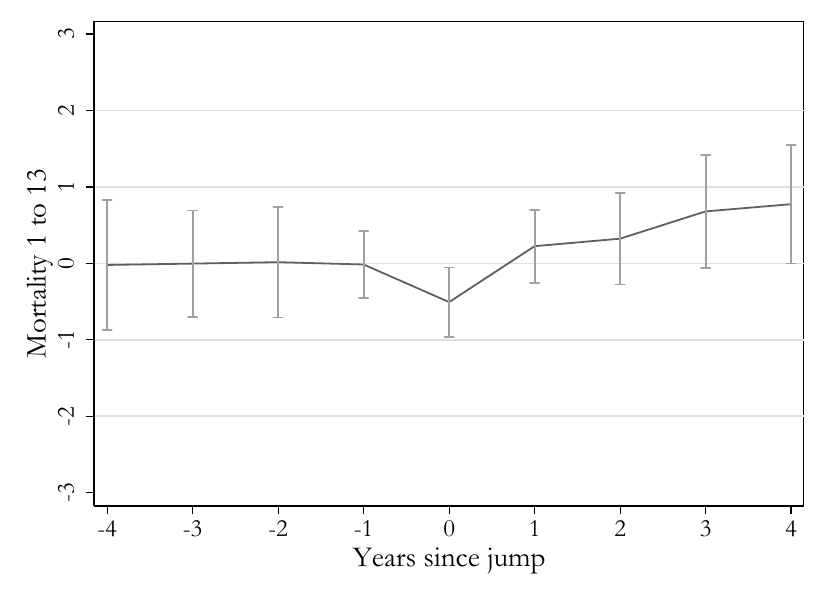}
        \caption{Ages 1 to 14}
    \end{subfigure}
    \begin{subfigure}[b]{0.45\linewidth}
        \centering
        \includegraphics[width=\linewidth]{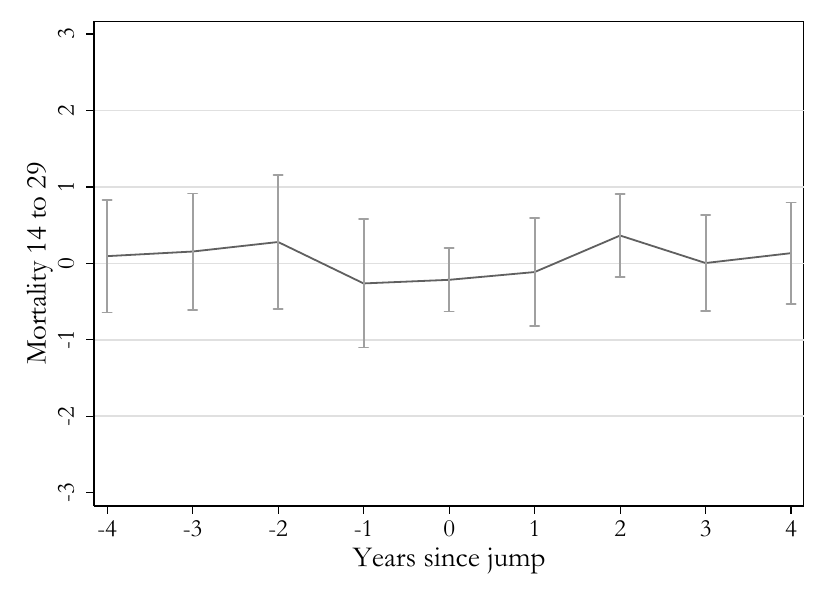}
        \caption{Ages 14 to 29}
    \end{subfigure}
    \begin{subfigure}[b]{0.45\linewidth}
        \centering
        \includegraphics[width=\linewidth]{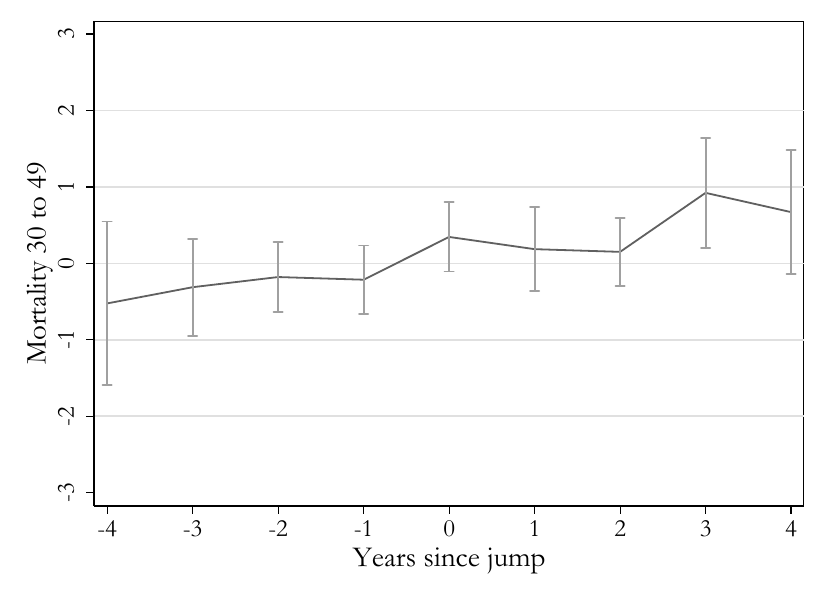}
        \caption{Ages 30 to 49}
    \end{subfigure}
    \begin{subfigure}[b]{0.45\linewidth}
        \centering
        \includegraphics[width=\linewidth]{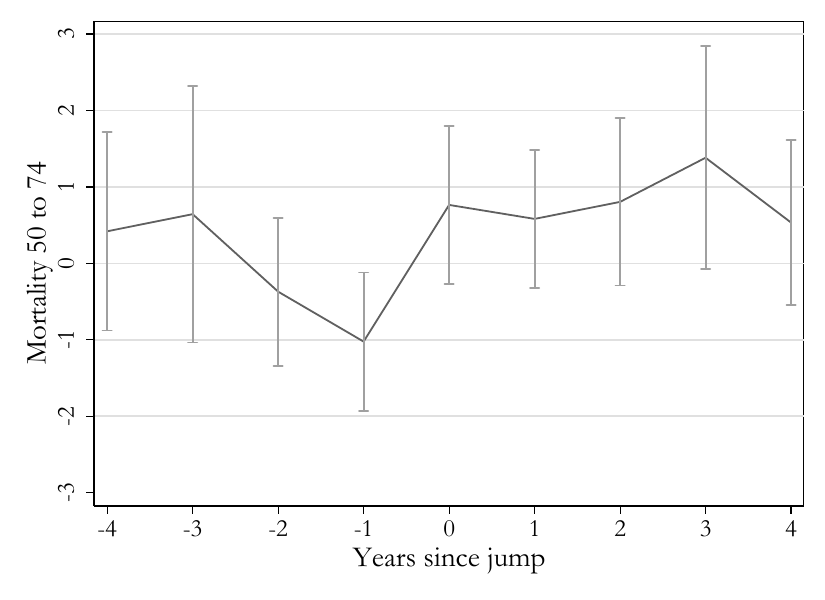}
        \caption{Ages 50 to 74}
    \end{subfigure}
    \begin{subfigure}[b]{0.45\linewidth}
        \centering
        \includegraphics[width=\linewidth]{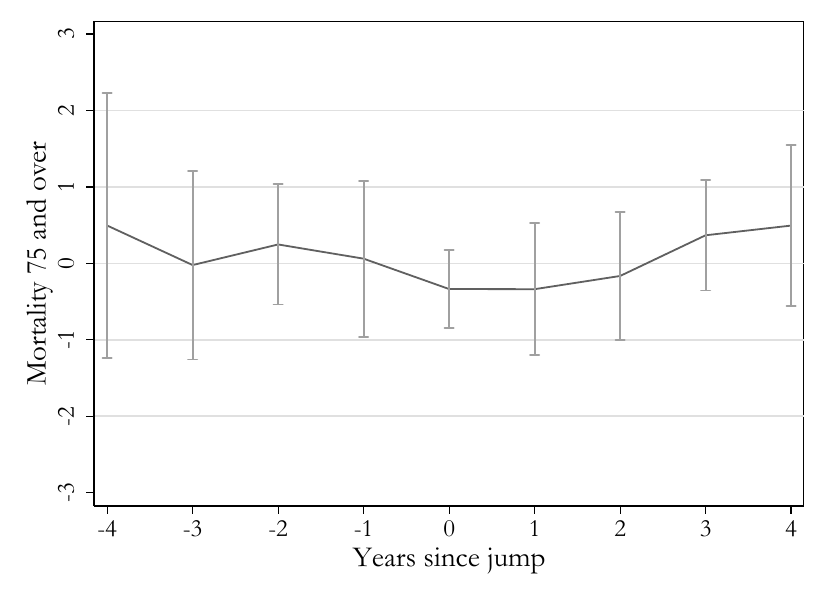}
        \caption{Ages 75-}
    \end{subfigure}
    \caption{Event Study: Effect by Age Group.}
    \label{fig:event-age}
\end{figure}

Figure \ref{fig:event-age} provides the corresponding event studies broken down by different age groups. For infant mortality, ages 14-29 and the oldest group (75+) the estimated effects are small and tend to be close to zero. On the other hand, children between ages 1 and 13, and individuals between 30 and 49 have elevated mortality in the years following the pollution shock. In general, the mortality effect appears to be building up over the consecutive years.

These estimates for specific age groups are summarised in Table \ref{tab:mainmort}, using the \citet{callaway2021difference} estimator. Again, the most pronounced effects are noted for the age groups 1-13, 30-49 and 50-74 whereas they are negligible for the age groups 0-1, 14-29 and 75+. Mortality increases by 0.82 deaths per 1,000 population in the 50 to 74 year old category. This corresponds to an increase of 10\% compared the treated group at baseline. Further, in the 30 to 49 year old bracket, we see a significant increase of 0.44 deaths per 1,000 population, an increase of 19.1\% compared to the baseline. 

\newpage 
{
\def\sym#1{\ifmmode^{#1}\else\(^{#1}\)\fi}
\begin{longtable}{l*{7}{c}}
\caption{Difference-in Differences Estimates of Mortality Effects of a Pollution Shock.}\label{tab:mainmort}\\
\toprule\endfirsthead\midrule\endhead\midrule\endfoot\endlastfoot
                  & \multicolumn{7}{c}{Outcome: Mortality by age bracket per 1000 baseline population}\\
                    &\multicolumn{1}{c}{(1)}&\multicolumn{1}{c}{(2)}&\multicolumn{1}{c}{(3)}&\multicolumn{1}{c}{(4)}&\multicolumn{1}{c}{(5)}&\multicolumn{1}{c}{(6)}&\multicolumn{1}{c}{(7)}\\
                    
                    &\multicolumn{1}{c}{All ages}&\multicolumn{1}{c}{$<1$}&\multicolumn{1}{c}{1 to 13}&\multicolumn{1}{c}{14 to 29}&\multicolumn{1}{c}{30 to 49}&\multicolumn{1}{c}{50 to 74}&\multicolumn{1}{c}{75 plus}\\
\midrule
ATT                 &      1.4103\sym{*}  &     -0.1104         &      0.2610         &      0.0267         &      0.4405\sym{**} &      0.8156\sym{*}  &     -0.0231         \\
                    &    (0.7932)         &    (0.1720)         &    (0.2272)         &    (0.2125)         &    (0.2045)         &    (0.4711)         &    (0.2705)         \\
\midrule
$N$                &          410           &        410             &          410           &        410             &          410           &         410            &       410              \\
\bottomrule
\multicolumn{8}{l}{\footnotesize DID estimates based on \citep{callaway2021difference} with never-treated units as comparison group.}\\
\multicolumn{8}{l}{\footnotesize Clustered standard errors in parentheses. \sym{*} \(p<.1\), \sym{**} \(p<.05\), \sym{***} \(p<.01\)}\\
\end{longtable}
}

\subsection{Robustness}

\begin{figure}[H]
       \centering
       \includegraphics[width=0.7\linewidth]{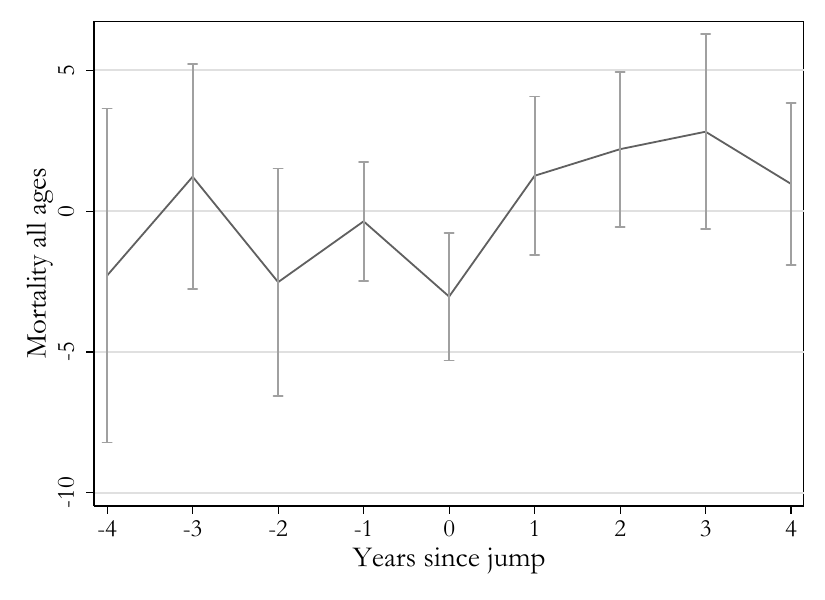}
\caption{Robustness: All-Age Mortality, controlling for water generated energy at baseline}
        \label{fig:mainmort_water}
         \end{figure}

\paragraph{Controlling for General Growth.} To assess whether our proxy also captures the effects of economic growth, we examine the impact of controlling for water turbine capacity on mortality rates. Results are presented in Figure \ref{fig:mainmort_water}. Our findings indicate that including water turbine capacity in the specification leaves estimates somewhat lower in precision, but similar in magnitude, suggesting that are main estimates are not biased by economic growth. This result is robust even when we employ a non-parametric TWFE specification shown in Table \ref{tab:waternonp}. Furthermore, the elasticity between the pollution proxy and water turbine capacity is 0.44 (within $R^2=0.74$), after controlling for parish and year fixed effects, indicating that any residual correlation between pollution and mortality, while controlling for water turbine capacity, is unlikely to be driven by economic growth. This likely explains the reduced precision of estimates, as they are now driven by residual correlation after controlling for hydropower.

\begin{figure}
       \centering
       \includegraphics[width=0.7\linewidth]{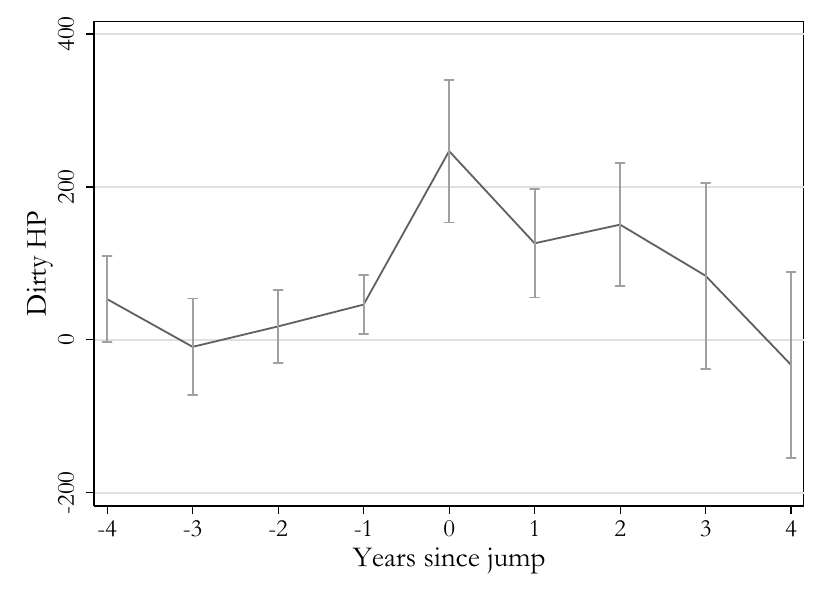}
\caption{First stage robustness: Impact on Dirty Horsepower, controlling for water generated energy at baseline}
        \label{fig:fstage_water}
         \end{figure}

When looking at ATT's by age group in Table \ref{tab:water}, we see that the magnitude of the 50 to 74 age group stays very similar (0.82 vs. 0.71)  but the 30 to 49 age group reduces to zero. This can potentially be driven by workplace accidents. the all-age mortality rate reduces to 0.72, which is about half of the baseline specification. Potentially, these estimates are biased, as we see a drop in mortality in the first year of the shock (period 0) in Figure \ref{fig:mainmort_water}.

{
\def\sym#1{\ifmmode^{#1}\else\(^{#1}\)\fi}
\begin{longtable}{l*{7}{c}}
\caption{Difference-in Differences Estimates: Controlling for Hydropower}\label{tab:water}\\
\toprule\endfirsthead\midrule\endhead\midrule\endfoot\endlastfoot
                   & \multicolumn{7}{c}{Outcome: Mortality by age bracket per 1000 baseline population}\\
                    &\multicolumn{1}{c}{(1)}&\multicolumn{1}{c}{(2)}&\multicolumn{1}{c}{(3)}&\multicolumn{1}{c}{(4)}&\multicolumn{1}{c}{(5)}&\multicolumn{1}{c}{(6)}&\multicolumn{1}{c}{(7)}\\
                    
                    &\multicolumn{1}{c}{All ages}&\multicolumn{1}{c}{$<1$}&\multicolumn{1}{c}{1 to 13}&\multicolumn{1}{c}{14 to 29}&\multicolumn{1}{c}{30 to 49}&\multicolumn{1}{c}{50 to 74}&\multicolumn{1}{c}{75 plus}\\
\midrule
ATT                 &      0.7215         &     -0.6377\sym{**} &      0.2534         &      0.1259         &      0.0156         &      0.7139         &      0.2505         \\
                    &    (1.0659)         &    (0.2683)         &    (0.4110)         &    (0.2888)         &    (0.3688)         &    (0.5797)         &    (0.6813)         \\
\midrule
Water turbine & Yes &Yes&Yes&Yes&Yes&Yes&Yes\\
$N$                &       402              &          401           &      401               &      401               &             401        &          401           &      401               \\
\bottomrule
\multicolumn{8}{l}{\footnotesize DID estimates based on \citep{callaway2021difference} with never-treated units as control group.}\\
\multicolumn{8}{l}{\footnotesize Clustered standard errors in parentheses. \sym{*} \(p<.1\), \sym{**} \(p<.05\), \sym{***} \(p<.01\)}\\
\multicolumn{8}{l}{\footnotesize Water turbine horsepower included as a control variable.}\\
\end{longtable}
}

\begin{figure}[H]
    \centering
    \begin{subfigure}[b]{0.45\linewidth}
        \centering
        \includegraphics[width=\linewidth]{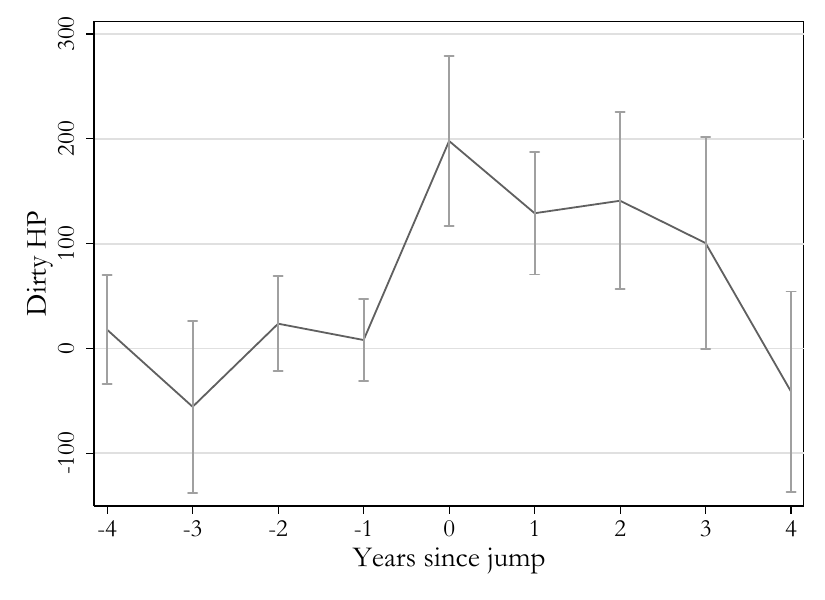}
        \caption{Impact on Pollution Proxy} \label{fig:fstage1920}
    \end{subfigure}
    \begin{subfigure}[b]{0.45\linewidth}
        \centering
        \includegraphics[width=\linewidth]{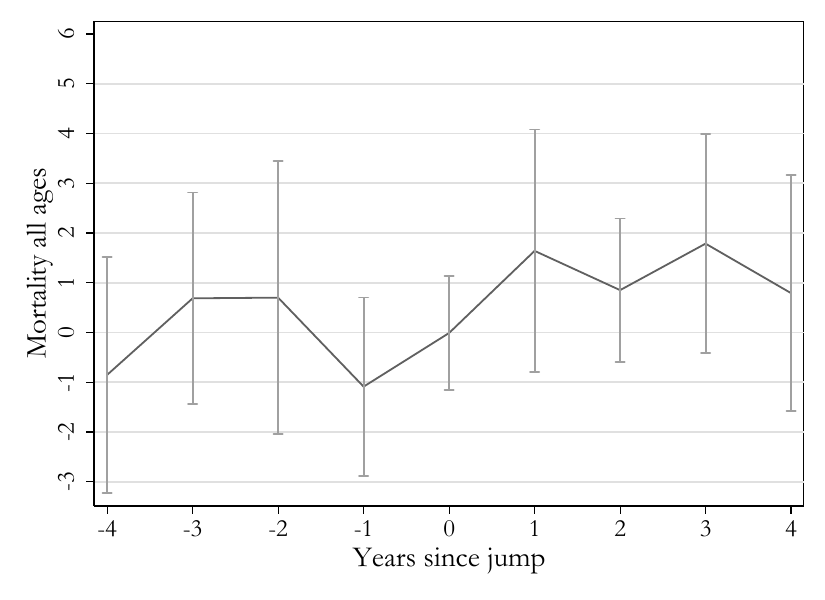}
        \caption{Impact on Mortality}\label{fig:mainmort1920}
    \end{subfigure}
    \caption{Robustness Check: Accounting for the 1921-22 Downturn.}
    \label{fig:robust1920}
\end{figure}

\paragraph{Accounting for 1921-1922 Economic Crisis.} As Figure \ref{fig:workforce} shows, the 1921-1922 Economic Crisis cam with a drastic reduction in workers across the country, which may impact our proxy for pollution in this period. In a robustness check, we freeze the pollution proxy in 1920, as the drastic reduction in workers may otherwise bias the treatment allocation. The first stage is shown in Figure \ref{fig:fstage1920} and the main event study plot is shown in Figure \ref{fig:mainmort1920}. All our main findings remain robust to this, as can be seen in the aggregate event study, as well as the age-group specific event studies in Figure \ref{fig:mortages1920} and the summary of age-specific ATT's in Table \ref{tab:mainmort1920}. 

{
\def\sym#1{\ifmmode^{#1}\else\(^{#1}\)\fi}
\begin{longtable}{l*{7}{c}}
\caption{Difference-in Differences Estimates: Accounting for the 1921-22 Downturn}\label{tab:water}\\
\toprule\endfirsthead\midrule\endhead\midrule\endfoot\endlastfoot
              & \multicolumn{7}{c}{Outcome: Mortality by age bracket per 1000 baseline population}\\
                    &\multicolumn{1}{c}{(1)}&\multicolumn{1}{c}{(2)}&\multicolumn{1}{c}{(3)}&\multicolumn{1}{c}{(4)}&\multicolumn{1}{c}{(5)}&\multicolumn{1}{c}{(6)}&\multicolumn{1}{c}{(7)}\\
                    
                    &\multicolumn{1}{c}{All ages}&\multicolumn{1}{c}{$<1$}&\multicolumn{1}{c}{1 to 13}&\multicolumn{1}{c}{14 to 29}&\multicolumn{1}{c}{30 to 49}&\multicolumn{1}{c}{50 to 74}&\multicolumn{1}{c}{75 plus}\\
\midrule
ATT                 &      1.0088         &     -0.0962         &     -0.0723         &      0.1054         &      0.5979\sym{*}  &      0.5626\sym{*}  &     -0.0886         \\
                    &    (0.6682)         &    (0.1752)         &    (0.2250)         &    (0.2226)         &    (0.3156)         &    (0.3128)         &    (0.2954)         \\
\midrule
$N$                &          411           &        411             &          411           &        411             &          411           &         411            &       411              \\
\bottomrule
\multicolumn{8}{l}{\footnotesize Clustered standard errors in parentheses.}\\
\multicolumn{8}{l}{\footnotesize \sym{*} \(p<.1\), \sym{**} \(p<.05\), \sym{***} \(p<.01\)}\\
\label{tab:mainmort1920}
\end{longtable}
}

\section{Conclusion}

This paper has examined the short-term impact of pollution on mortality in urban areas during a critical period of industrialisation in Sweden. By leveraging unique historical data and employing a novel empirical strategy, we have quantified the detrimental effects of pollution driven by the rapid industrial growth of the early \nth{20} century. Our study contributes to the literature on the historical determinants of health by addressing several key gaps, particularly the lack of individual-level mortality data and the limited focus on specific pollutants and their direct effects on mortality.

Through a combination of innovative data sources---ranging from detailed mortality records to local energy production data---we constructed a credible measure of pollution and identified its direct impact on mortality. Using a two-way fixed-effects model and exploiting natural variations in hydropower availability across towns, we isolated the effects of pollution from other socio-economic changes driven by industrial growth. This allowed us to provide a more accurate estimate of the pollution-mortality relationship and ensures that our results are not confounded by broader economic factors.

Our findings indicate that local pollution shocks resulted in a gradual increase in mortality rates, with a significant effect size observed after four years. The overall average treatment effect (ATT) corresponds to approximately 1.4 additional deaths per 1,000 population, a substantial increase of about 8 percent over baseline mortality. The effect was most pronounced among certain age groups, with children (ages 1-13), individuals aged 30-49, and those aged 50-74 experiencing notable increases in mortality rates. In particular, mortality among individuals aged 30-49 increased by 19.1 percent, while in the 50-74 age group, it increased by 10 percent relative to the baseline. The pollution shock persisted for around four years, with the effects gradually intensifying over time.

These findings have important implications for our understanding of the historical role of environmental factors in shaping public health outcomes. While the industrial revolution brought substantial economic benefits, it also introduced severe environmental costs that had direct and lasting consequences on public health. The historical context provided by this study can inform contemporary debates on the impact of pollution in developing economies today, many of which are experiencing pollution levels reminiscent of those seen during early industrialization in high-income countries.

In conclusion, this paper underscores the importance of considering environmental factors, such as pollution, alongside economic and technological advances when evaluating the determinants of life expectancy. It also highlights the value of historical data and methods, particularly the novel use of town-level industrial and energy data, in advancing our understanding of long-term health outcomes. This study provides a foundation for further research into the complex interplay between industrial growth, environmental degradation, and public health.

\newpage
\setcitestyle{authoryear}
\bibliography{Literature} 

\newpage
\renewcommand\thesection{A}
\renewcommand\thetable{\thesection\arabic{table}} \setcounter{table}{0}
\renewcommand\thefigure{\thesection\arabic{figure}} \setcounter{figure}{0}

\section{Appendix}
\subsection{Descriptive statistics}

   \begin{figure}[H]
       \centering
        \includegraphics[width=0.7\linewidth]{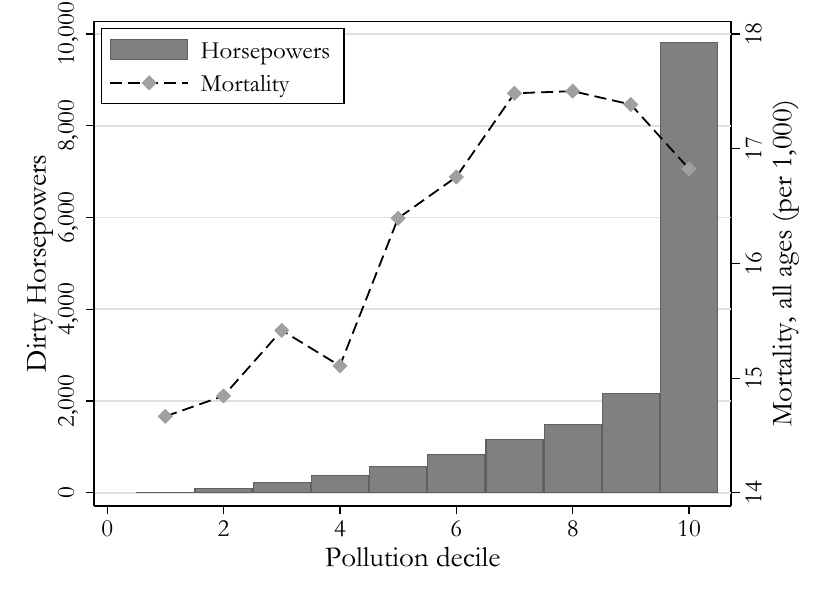}
\caption{Relationship between Mortality Rates and Pollution Exposure Deciles}
        \label{fig:MortGraph2}
         \end{figure}

  \begin{figure}
        \centering
        \includegraphics[width=0.4\linewidth]{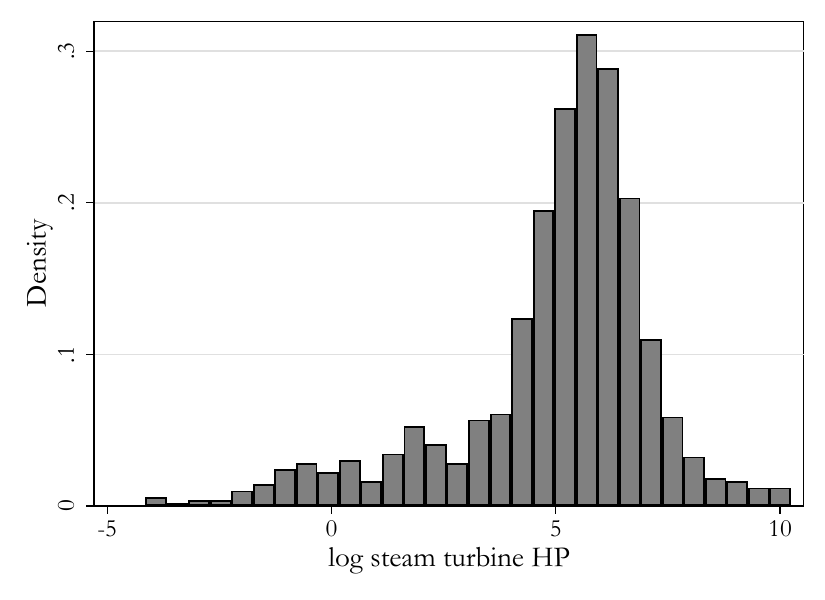}
              \includegraphics[width=0.4\linewidth]{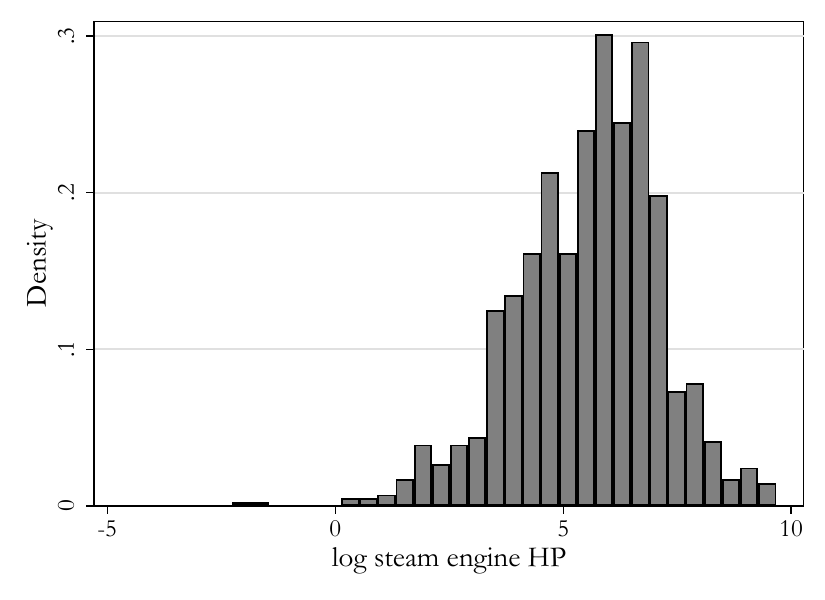}
                    \includegraphics[width=0.4\linewidth]{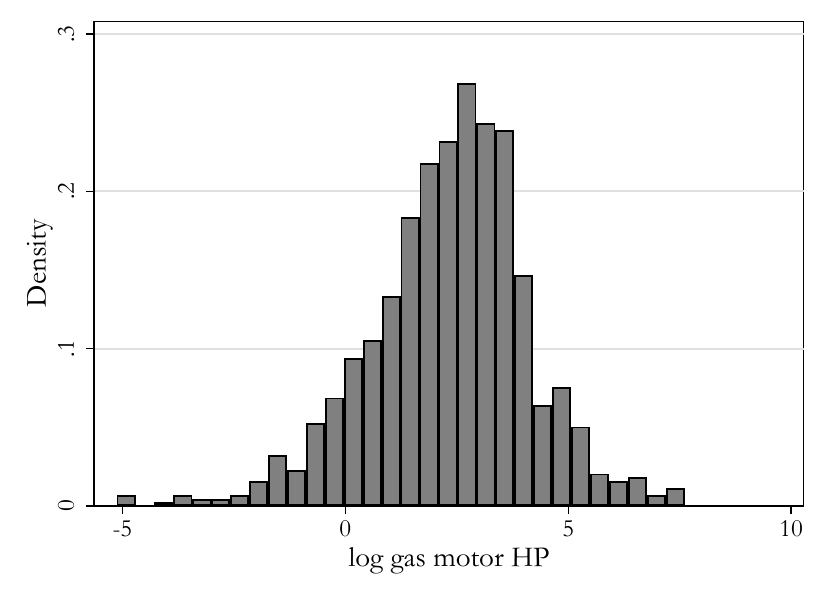}
                          \includegraphics[width=0.4\linewidth]{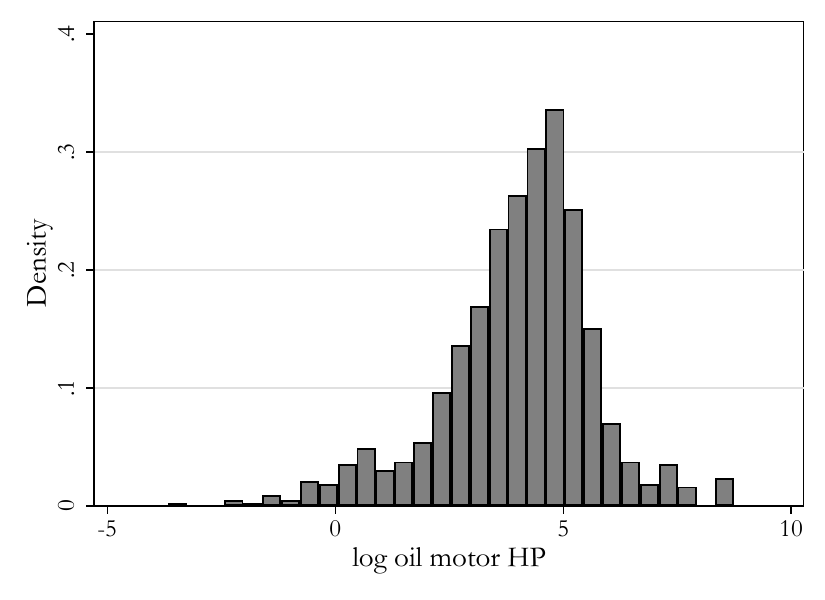}  
                                  \includegraphics[width=0.4\linewidth]{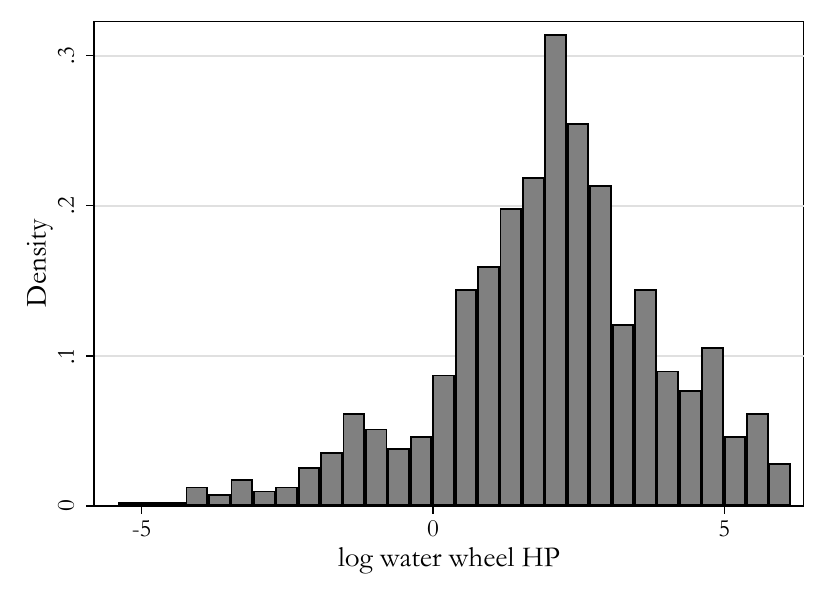} 
                                      \includegraphics[width=0.4\linewidth]{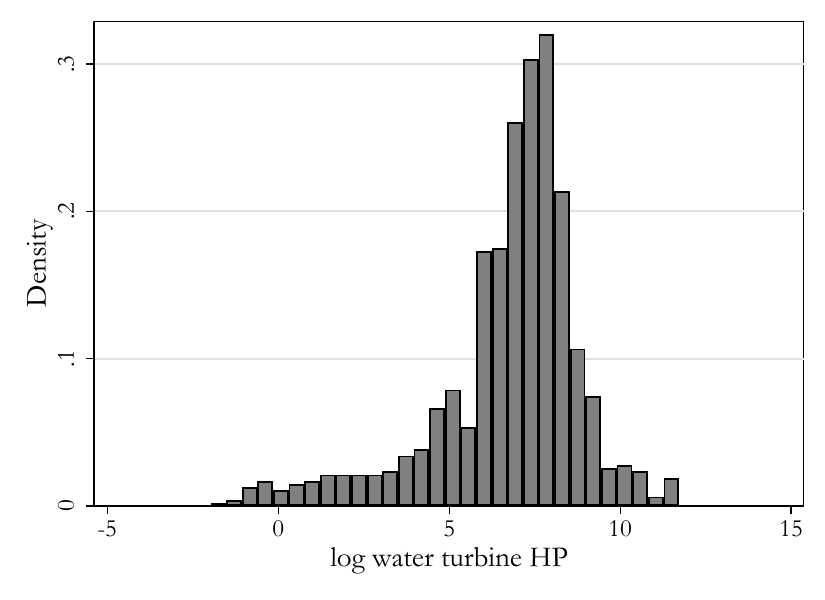}    
        \caption{log of energy by source}
        \label{fig:individual_dirty}
    \end{figure}
    
     \begin{figure}
        \centering
        \includegraphics[width=0.4\linewidth]{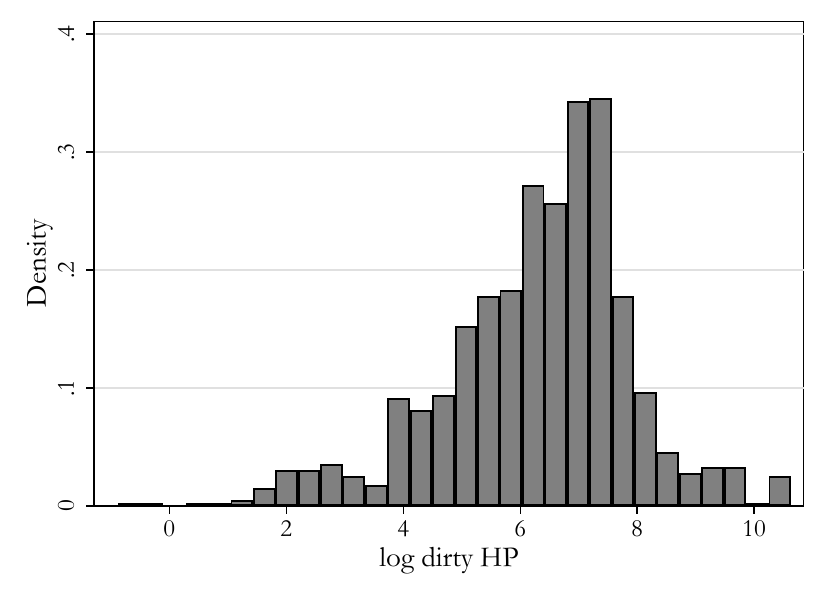}
              \includegraphics[width=0.4\linewidth]{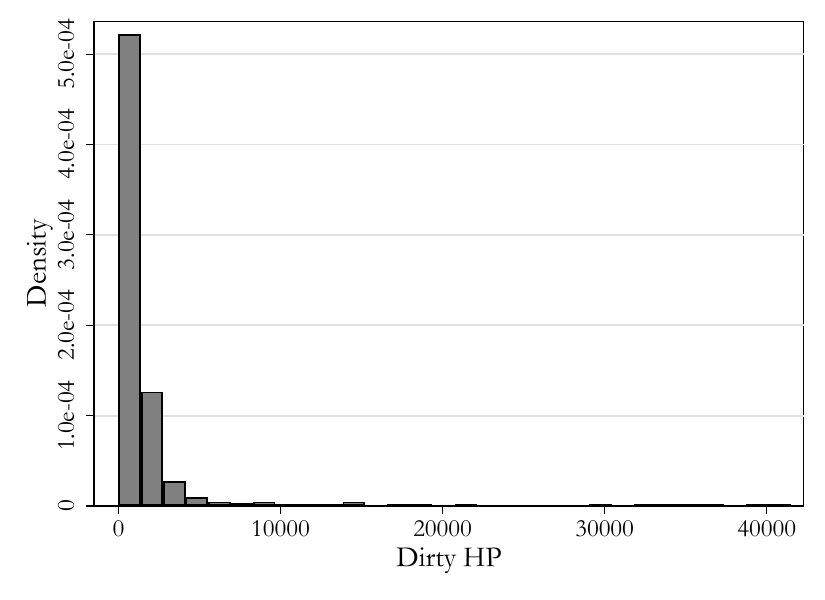}
        \caption{Dirty energy proxy, log and raw}
        \label{fig:sumdirty_dist}
    \end{figure}

\begin{figure}
       \centering
       \includegraphics[width=0.4\linewidth]{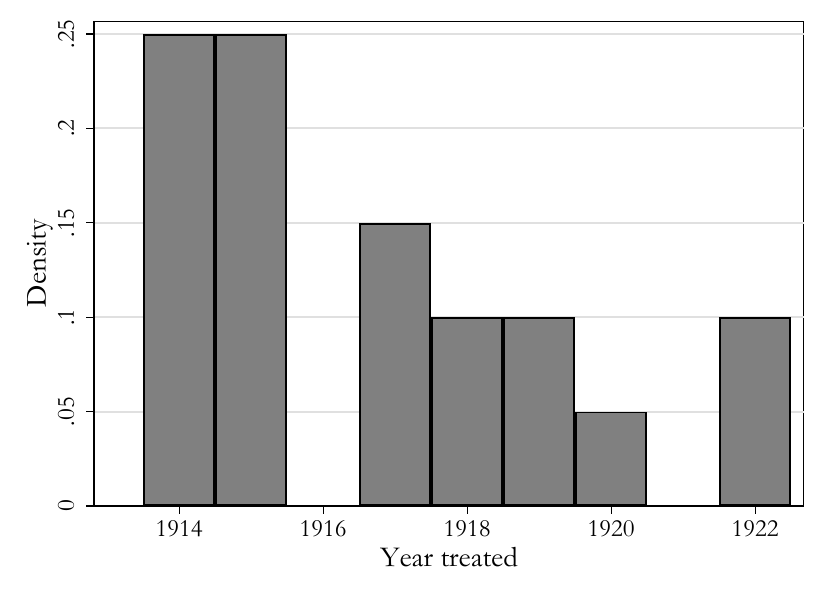}
       \includegraphics[width=0.4\linewidth]{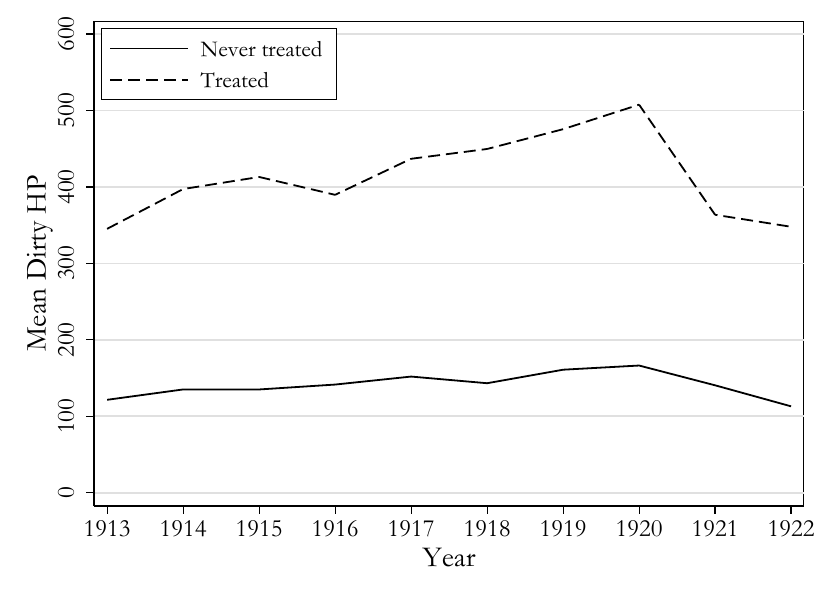}
        \includegraphics[width=0.4\linewidth]{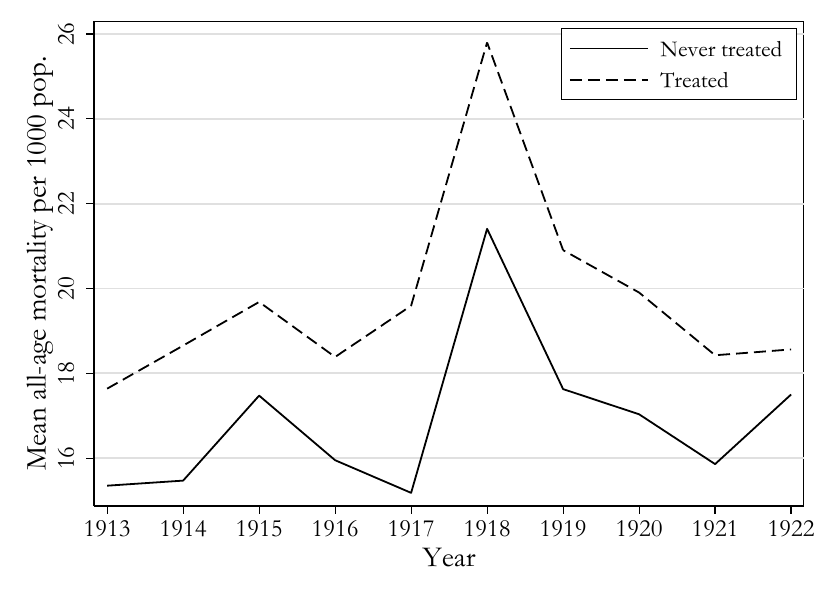}
\caption{Treatment 4th and lower to 5th and above deciles of sum dirty}
        \label{fig:enter-label}
         \end{figure}

{
\def\sym#1{\ifmmode^{#1}\else\(^{#1}\)\fi}
\begin{longtable}{l*{1}{cccccc}}
\caption{Summary Statistics: full sample}\\
\toprule\endfirsthead\midrule\endhead\midrule\endfoot\endlastfoot
                    &\multicolumn{6}{c}{}                                                         \\
                    &        Obs.&        Mean&        Med.&        S.D.&        Min.&        Max.\\
\midrule
Mortality all ages  &       1,070&      16.335&      14.462&        9.00&         3.1&        85.7\\
Mortality under 1   &       1,070&       1.456&       1.270&        1.04&         0&         9.6\\
Mortality 1 to 13   &       1,070&       1.364&       1.113&        1.12&         0&        10.9\\
Mortality 14 to 29  &       1,070&       1.951&       1.601&        1.50&         0&        17.6\\
Mortality 30 to 49  &       1,070&       2.173&       1.890&        1.40&         0&        14.3\\
Mortality 50 to 74  &       1,070&       5.172&       4.504&        3.05&         0.6&        27.2\\
Mortality 75 and over&       1,070&       4.218&       3.395&        3.15&         0&        25.4\\
Deaths all ages     &       1,100&     208.336&      88.000&      584.22&         7&      6796\\
Deaths under 1      &       1,100&      21.007&       8&       55.96&         0&       632\\
Deaths 1 to 13      &       1,100&      18.822&       7&       52.51&         0&       546\\
Deaths 14 to 29     &       1,100&      27.737&      11&       87.33&         0&      1430\\
Deaths 30 to 49     &       1,100&      32.679&      12&      109.13&         0&      1580\\
Deaths 50 to 74     &       1,100&      64.595&      27&      187.48&         1&      1914\\
Deaths 75 and over  &       1,100&      43.496&      21&      105.61&         0&      1128\\
Population          &       1,100&  15,540.250&   5,950&   43,997.99&         0&    422042\\
Dirty HP            &       1,027&   1,682.773&     696.485&    4,107.65&         0.4&     41502.4\\
log dirty HP        &       1,027&       6.298&       6.546&        1.64&        -0.9&        10.6\\
Coastal             &       1,100&       0.527&       1&        0.50&         0&         1\\
\bottomrule
\end{longtable}
\label{sum_full}
}

\subsection{Two-way fixed effects results}

{
\def\sym#1{\ifmmode^{#1}\else\(^{#1}\)\fi}
\begin{longtable}{l*{1}{cccccc}}
\caption{Summary Statistics: TWFE sample}\\
\toprule\endfirsthead\midrule\endhead\midrule\endfoot\endlastfoot
                    &\multicolumn{6}{c}{}                                                         \\
                    &        Obs.&        Mean&        Med.&        S.D.&        Min.&        Max.\\
\midrule
Mortality all ages  &         901&      16.236&      14.555&        8.24&         3.2&        83.0\\
Mortality under 1   &         901&       1.414&       1.237&        1.00&         0.0&         9.6\\
Mortality 1 to 13   &         901&       1.337&       1.109&        1.07&         0.0&        10.1\\
Mortality 14 to 29  &         901&       1.950&       1.616&        1.39&         0.0&        10.5\\
Mortality 30 to 49  &         901&       2.157&       1.903&        1.30&         0.0&        12.4\\
Mortality 50 to 74  &         901&       5.153&       4.523&        2.92&         0.9&        27.2\\
Mortality 75 and over&         901&       4.224&       3.445&        3.00&         0.0&        25.4\\
Deaths all ages     &         901&     220.926&      90.000&      614.49&         9.0&      6796.0\\
Deaths under 1      &         901&      21.576&       8.000&       57.03&         0.0&       632.0\\
Deaths 1 to 13      &         901&      19.564&       7.000&       55.37&         0.0&       546.0\\
Deaths 14 to 29     &         901&      29.614&      11.000&       93.39&         0.0&      1430.0\\
Deaths 30 to 49     &         901&      34.779&      12.000&      114.78&         0.0&      1580.0\\
Deaths 50 to 74     &         901&      68.868&      28.000&      196.68&         2.0&      1914.0\\
Deaths 75 and over  &         901&      46.525&      22.000&      111.89&         0.0&      1128.0\\
Population    &         901&  16,590.114&   6,013.000&   46,426.17&       598.0&    422042.0\\
Dirty HP            &         897&   1,634.529&     698.590&    3,991.59&         0.7&     40890.2\\
log dirty HP        &         897&       6.305&       6.549&        1.61&        -0.4&        10.6\\
Coastal             &         901&       0.552&       1.000&        0.50&         0.0&         1.0\\
\bottomrule
\end{longtable}
\label{sum_twfe}
}
    {
\def\sym#1{\ifmmode^{#1}\else\(^{#1}\)\fi}
\begin{longtable}{l*{7}{c}}
\caption{OLS with parish and year fixed effects, SE clustered at parish}\\
\toprule\endfirsthead\midrule\endhead\midrule\endfoot\endlastfoot
                         &\multicolumn{1}{c}{(1)}&\multicolumn{1}{c}{(2)}&\multicolumn{1}{c}{(3)}&\multicolumn{1}{c}{(4)}&\multicolumn{1}{c}{(5)}&\multicolumn{1}{c}{(6)}&\multicolumn{1}{c}{(7)}\\
                         &\multicolumn{1}{c}{All ages}&\multicolumn{1}{c}{Under 1}&\multicolumn{1}{c}{1 to 13}&\multicolumn{1}{c}{14 to 29}&\multicolumn{1}{c}{30 to 49}&\multicolumn{1}{c}{50 to 74}&\multicolumn{1}{c}{75 and over}\\
\midrule
 2nd Decile Dirty$_{t-1}$&    0.1781         &      0.1426         &     -0.0073         &      0.1597         &      0.3071         &      0.1247         &     -0.5489         \\
                         &    (0.8639)         &    (0.2216)         &    (0.1366)         &    (0.2522)         &    (0.2011)         &    (0.4496)         &    (0.3474)         \\
3rd decile Dirty$_{t-1}$&       0.7502         &      0.2618         &      0.1041         &      0.3185         &      0.1486         &     -0.0763         &     -0.0065         \\
                         &    (0.9370)         &    (0.2993)         &    (0.2183)         &    (0.3759)         &    (0.3099)         &    (0.4085)         &    (0.4995)         \\
4th decile Dirty$_{t-1}$&       0.4404         &      0.3169         &      0.0017         &      0.1634         &      0.0240         &      0.2448         &     -0.3104         \\
                         &    (0.9827)         &    (0.3385)         &    (0.2231)         &    (0.3692)         &    (0.3300)         &    (0.4245)         &    (0.5640)         \\
5th decile Dirty$_{t-1}$&      1.7274\sym{*}  &      0.5538\sym{*}  &      0.4682\sym{*}  &      0.0253         &      0.2892         &      0.5401         &     -0.1492         \\
                         &    (1.0282)         &    (0.3265)         &    (0.2388)         &    (0.3431)         &    (0.3317)         &    (0.5015)         &    (0.5804)         \\
6th decile Dirty$_{t-1}$&       2.0871\sym{*}  &      0.7151\sym{**} &      0.7652\sym{***}&     -0.0966         &      0.3921         &      0.6932         &     -0.3818         \\
                         &    (1.0589)         &    (0.3391)         &    (0.2902)         &    (0.3517)         &    (0.3644)         &    (0.5142)         &    (0.6867)         \\
7th decile Dirty$_{t-1}$&      2.8163\sym{***}&      0.8017\sym{**} &      0.7613\sym{**} &      0.0439         &      0.5043         &      0.9686\sym{*}  &     -0.2636         \\
                         &    (1.0648)         &    (0.3471)         &    (0.2931)         &    (0.3657)         &    (0.3711)         &    (0.5438)         &    (0.6716)         \\
8th decile Dirty$_{t-1}$&       2.8350\sym{**} &      0.6610\sym{*}  &      0.6742\sym{**} &      0.1792         &      0.4381         &      1.0856\sym{*}  &     -0.2032         \\
                         &    (1.1035)         &    (0.3556)         &    (0.3099)         &    (0.3904)         &    (0.3806)         &    (0.5581)         &    (0.6696)         \\
9th decile Dirty$_{t-1}$&       2.7187\sym{**} &      0.7447\sym{**} &      0.6942\sym{**} &      0.2561         &      0.4897         &      0.8472         &     -0.3131         \\
                         &    (1.2115)         &    (0.3621)         &    (0.3195)         &    (0.4681)         &    (0.4235)         &    (0.6030)         &    (0.7001)         \\
10th decile Dirty$_{t-1}$&       2.1580\sym{*}  &      0.6602\sym{*}  &      0.6246\sym{*}  &     -0.0887         &      0.5269         &      0.8505         &     -0.4155         \\
                         &    (1.2600)         &    (0.3618)         &    (0.3435)         &    (0.4748)         &    (0.4344)         &    (0.6248)         &    (0.7073)         \\
Constant                 &     14.6667\sym{***}&      0.9290\sym{***}&      0.9304\sym{***}&      1.8512\sym{***}&      1.8453\sym{***}&      4.6278\sym{***}&      4.4829\sym{***}\\
                         &    (0.8446)         &    (0.2770)         &    (0.2061)         &    (0.2940)         &    (0.2786)         &    (0.4131)         &    (0.4985)         \\
\midrule
   $N$              &         901         &         901         &         901         &         901         &         901         &         901         &         901         \\
  $R^2$                     &      0.9119         &      0.6686         &      0.6373         &      0.6725         &      0.6371         &      0.8224         &      0.8639         \\
\bottomrule
\multicolumn{8}{l}{\footnotesize Clustered standard errors at parish level in parentheses}\\
\multicolumn{8}{l}{\footnotesize \sym{*} \(p<.1\), \sym{**} \(p<.05\), \sym{***} \(p<.01\)}\\
 \caption{TWFE results of mortality per 1,000 inhabitants (total parish population measured at baseline) on dirty engine capacity in quantiles.}
\end{longtable}
\label{tab:TWFE}
}

\subsection{Water-driven power}

{
\def\sym#1{\ifmmode^{#1}\else\(^{#1}\)\fi}
\begin{longtable}{l*{7}{c}}
\toprule\endfirsthead\midrule\endhead\midrule\endfoot\endlastfoot
                            &\multicolumn{1}{c}{(1)}&\multicolumn{1}{c}{(2)}&\multicolumn{1}{c}{(3)}&\multicolumn{1}{c}{(4)}&\multicolumn{1}{c}{(5)}&\multicolumn{1}{c}{(6)}&\multicolumn{1}{c}{(7)}\\
                         &\multicolumn{1}{c}{All ages}&\multicolumn{1}{c}{Under 1}&\multicolumn{1}{c}{1 to 13}&\multicolumn{1}{c}{14 to 29}&\multicolumn{1}{c}{30 to 49}&\multicolumn{1}{c}{50 to 74}&\multicolumn{1}{c}{75 and over}\\
\midrule
 2nd Decile Dirty$_{t-1}$&      0.2735         &      0.3559         &     -0.0661         &      0.0430         &      0.1762         &      0.4711         &     -0.7065         \\
                         &    (1.0377)         &    (0.2857)         &    (0.1809)         &    (0.2648)         &    (0.2023)         &    (0.5051)         &    (0.4659)         \\
 3rd Decile Dirty$_{t-1}$&      0.6434         &      0.5985         &      0.1782         &      0.1209         &     -0.0828         &      0.1798         &     -0.3511         \\
                         &    (1.1582)         &    (0.3822)         &    (0.2699)         &    (0.4709)         &    (0.4102)         &    (0.5693)         &    (0.5533)         \\
 4th Decile Dirty$_{t-1}$&      0.3870         &      0.7398\sym{*}  &      0.1500         &     -0.0918         &     -0.1831         &      0.3512         &     -0.5791         \\
                         &    (1.2989)         &    (0.4312)         &    (0.2749)         &    (0.4335)         &    (0.4428)         &    (0.5663)         &    (0.6549)         \\
 5th Decile Dirty$_{t-1}$&      1.9486         &      1.0280\sym{**} &      0.6941\sym{**} &     -0.3133         &      0.1045         &      0.6126         &     -0.1773         \\
                         &    (1.4123)         &    (0.4407)         &    (0.3434)         &    (0.4250)         &    (0.4683)         &    (0.6271)         &    (0.7095)         \\
6th Decile Dirty$_{t-1}$&      2.2436         &      1.1901\sym{**} &      0.9734\sym{**} &     -0.4641         &      0.1792         &      0.7338         &     -0.3687         \\
                         &    (1.4711)         &    (0.4639)         &    (0.3713)         &    (0.4575)         &    (0.5143)         &    (0.6535)         &    (0.7777)         \\
7th Decile Dirty$_{t-1}$&      2.8418\sym{*}  &      1.2668\sym{***}&      0.9579\sym{**} &     -0.3181         &      0.2222         &      1.0152         &     -0.3022         \\
                         &    (1.5025)         &    (0.4763)         &    (0.3790)         &    (0.4849)         &    (0.5226)         &    (0.6930)         &    (0.7703)         \\
8th Decile Dirty$_{t-1}$&      3.0249\sym{*}  &      1.1427\sym{**} &      0.9246\sym{**} &     -0.1778         &      0.1541         &      1.1523         &     -0.1710         \\
                         &    (1.5836)         &    (0.4901)         &    (0.4009)         &    (0.5204)         &    (0.5482)         &    (0.7126)         &    (0.7751)         \\
9th Decile Dirty$_{t-1}$&      2.9840\sym{*}  &      1.2331\sym{**} &      0.9628\sym{**} &     -0.0705         &      0.1650         &      0.9322         &     -0.2386         \\
                         &    (1.7177)         &    (0.5008)         &    (0.4205)         &    (0.6236)         &    (0.6074)         &    (0.7553)         &    (0.8002)         \\
10th Decile Dirty$_{t-1}$&      2.4218         &      1.2032\sym{**} &      0.9203\sym{**} &     -0.3974         &      0.1678         &      0.9265         &     -0.3984         \\
                         &    (1.7748)         &    (0.5131)         &    (0.4408)         &    (0.6421)         &    (0.6251)         &    (0.7773)         &    (0.8229)         \\
2nd Decile Hydro$_{t-1}$&     -0.8611         &      0.2876         &      0.4963\sym{***}&      0.3032         &      0.2096         &     -0.8505         &     -1.3073\sym{*}  \\
                         &    (1.3537)         &    (0.2601)         &    (0.1789)         &    (0.2231)         &    (0.3062)         &    (0.9484)         &    (0.7532)         \\
3rd Decile Hydro$_{t-1}$&     -0.7870         &     -0.1982         &      0.4808\sym{*}  &      0.3511         &      0.3225         &     -1.2019         &     -0.5413         \\
                         &    (1.3186)         &    (0.2854)         &    (0.2747)         &    (0.2602)         &    (0.3879)         &    (0.8980)         &    (0.8003)         \\
4th Decile Hydro$_{t-1}$&     -0.3518         &     -0.2008         &      0.1530         &      0.4502         &      0.4440         &     -0.8081         &     -0.3902         \\
                         &    (1.3338)         &    (0.3188)         &    (0.2470)         &    (0.2835)         &    (0.3597)         &    (0.9349)         &    (0.7809)         \\
5th Decile Hydro$_{t-1}$&     -0.4819         &     -0.2716         &      0.2986         &      0.4519         &      0.3082         &     -0.6007         &     -0.6684         \\
                         &    (1.3310)         &    (0.3468)         &    (0.2905)         &    (0.3044)         &    (0.3919)         &    (0.8873)         &    (0.8046)         \\
6th Decile Hydro$_{t-1}$&     -1.1777         &     -0.3091         &      0.0323         &      0.5951\sym{*}  &      0.3341         &     -0.7869         &     -1.0431         \\
                         &    (1.3783)         &    (0.3344)         &    (0.2910)         &    (0.3230)         &    (0.3931)         &    (0.9051)         &    (0.8020)         \\
7th Decile Hydro$_{t-1}$&     -0.6756         &     -0.3003         &      0.2381         &      0.5892\sym{*}  &      0.3145         &     -0.5879         &     -0.9292         \\
                         &    (1.3989)         &    (0.3475)         &    (0.3119)         &    (0.3370)         &    (0.4200)         &    (0.9108)         &    (0.8229)         \\
8th Decile Hydro$_{t-1}$&     -0.3711         &     -0.2408         &      0.2921         &      0.5759         &      0.4932         &     -0.6949         &     -0.7967         \\
                         &    (1.4373)         &    (0.3595)         &    (0.3229)         &    (0.3587)         &    (0.4286)         &    (0.9206)         &    (0.8607)         \\
9th Decile Hydro$_{t-1}$&     -0.9559         &     -0.2672         &      0.1760         &      0.5138         &      0.4851         &     -0.8355         &     -1.0281         \\
                         &    (1.4690)         &    (0.3672)         &    (0.3306)         &    (0.3831)         &    (0.4493)         &    (0.9401)         &    (0.8662)         \\
10th Decile Hydro$_{t-1}$&     -0.7580         &     -0.4097         &      0.1441         &      0.5212         &      0.5469         &     -0.7185         &     -0.8420         \\
                         &    (1.5287)         &    (0.3849)         &    (0.3594)         &    (0.4173)         &    (0.4672)         &    (0.9649)         &    (0.8902)         \\
Constant                 &     15.2449\sym{***}&      0.7487\sym{**} &      0.5268\sym{***}&      1.6465\sym{***}&      1.6959\sym{***}&      5.2750\sym{***}&      5.3521\sym{***}\\
                         &    (1.3419)         &    (0.3413)         &    (0.2003)         &    (0.3185)         &    (0.3752)         &    (0.8044)         &    (0.6703)         \\
\midrule
  $N$             &         901         &         901         &         901         &         901         &         901         &         901         &         901         \\
 $R^2$                      &      0.9127         &      0.6758         &      0.6460         &      0.6739         &      0.6390         &      0.8247         &      0.8687         \\
\bottomrule
\multicolumn{8}{l}{\footnotesize Clustered standard errors at parish level in parentheses}\\
\multicolumn{8}{l}{\footnotesize \sym{*} \(p<.1\), \sym{**} \(p<.05\), \sym{***} \(p<.01\)}\\
\caption{OLS with parish and year fixed effects, SE clustered at parish}\label{tab:robust1}\\
\label{tab:waternonp}
\end{longtable}
}

\begin{figure}
       \centering
       \includegraphics[width=0.7\linewidth]{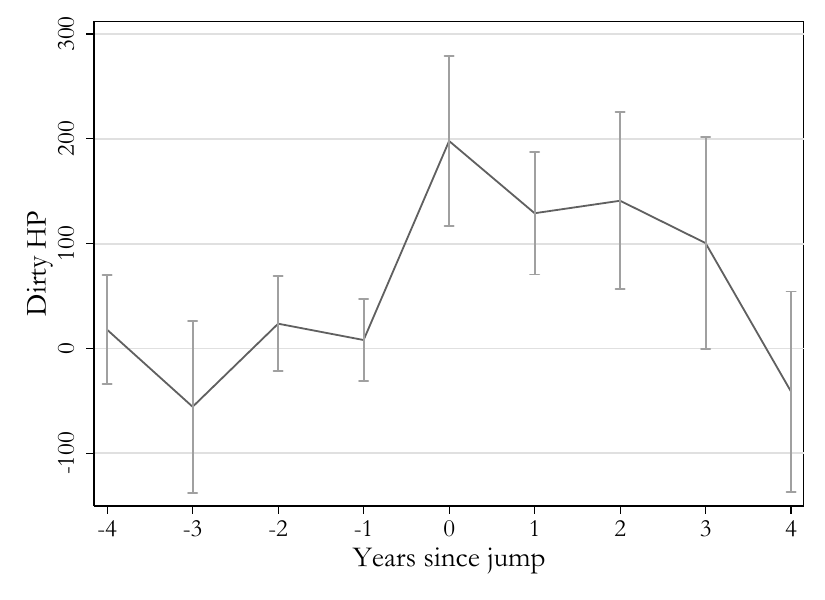}
\caption{First stage: Impact on Dirty Horsepowers while accounting for Economic Crisis of 1921.}
        \label{fig:fstage}
         \end{figure}

\begin{figure}[H]
       \centering
       \includegraphics[width=0.7\linewidth]{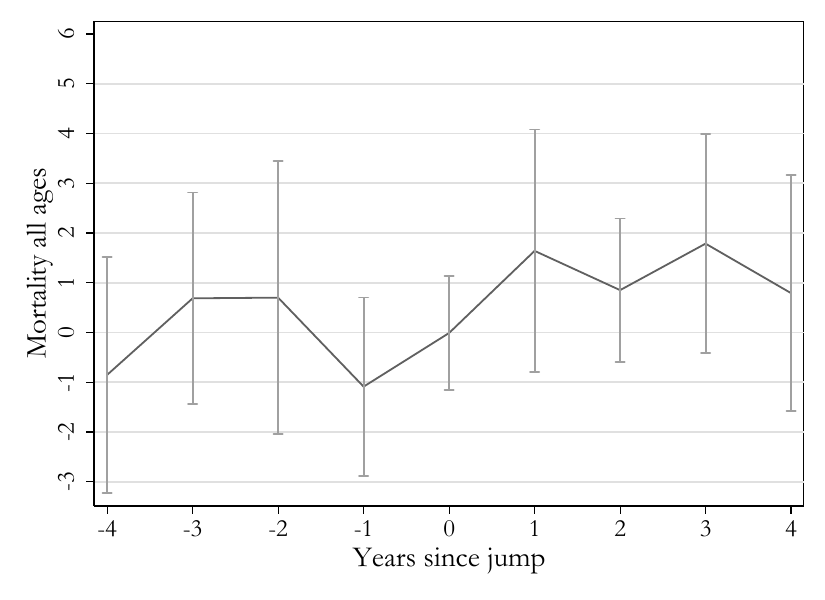}
\caption{Main Results: All-Age Mortality while accounting for Economic Crisis of 1921.}
        \label{fig:mainmort}
         \end{figure}

\begin{figure}[H]
    \centering
    \begin{subfigure}[b]{0.45\linewidth}
        \centering
        \includegraphics[width=\linewidth]{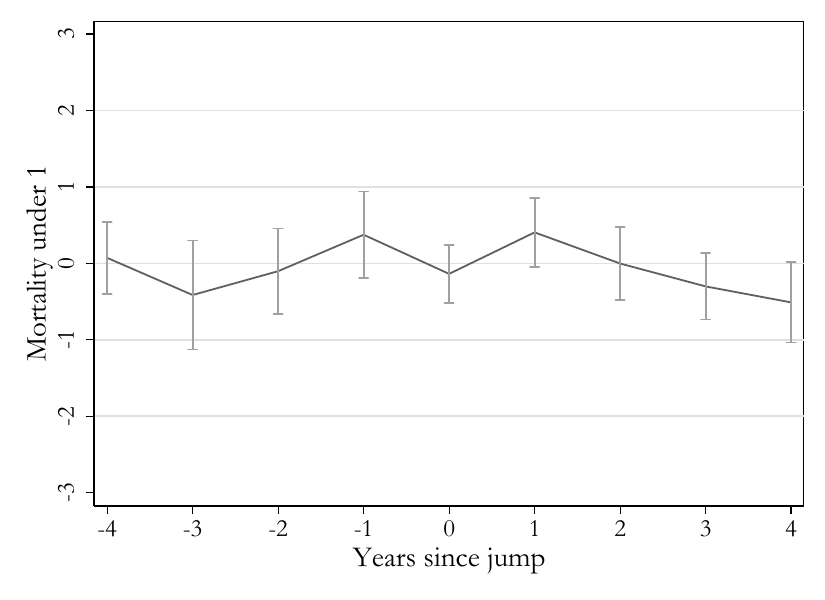}
        \caption{Infant Mortality}
    \end{subfigure}
    \begin{subfigure}[b]{0.45\linewidth}
        \centering
        \includegraphics[width=\linewidth]{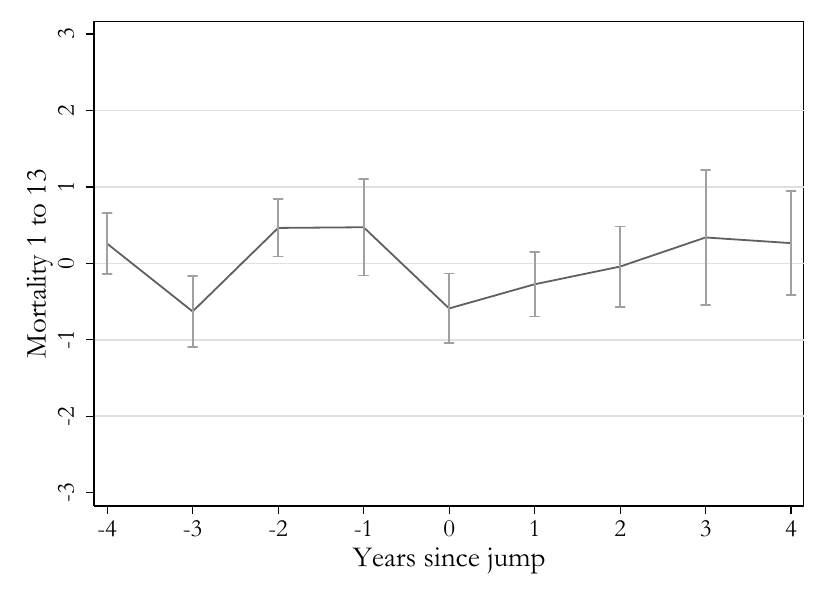}
        \caption{Ages 1 to 14}
    \end{subfigure}
    \begin{subfigure}[b]{0.45\linewidth}
        \centering
        \includegraphics[width=\linewidth]{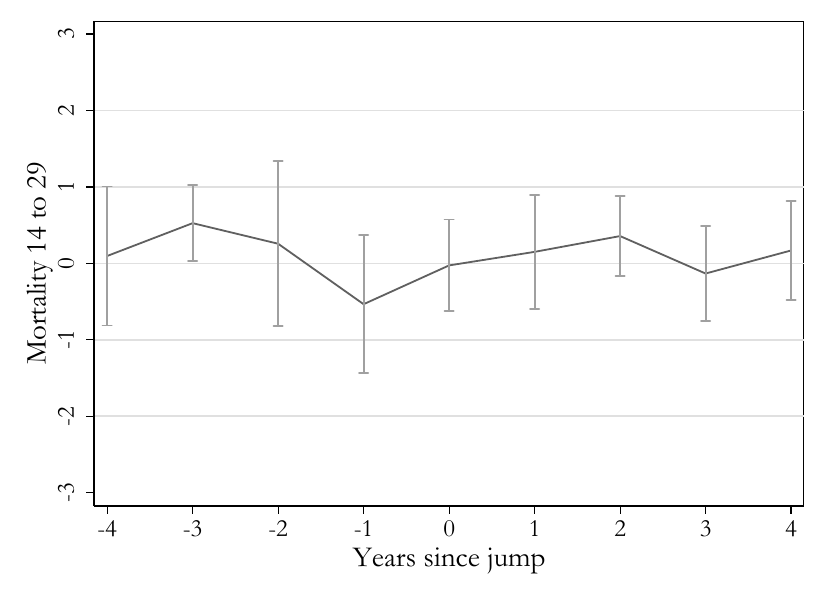}
        \caption{Ages 14 to 29}
    \end{subfigure}
    \begin{subfigure}[b]{0.45\linewidth}
        \centering
        \includegraphics[width=\linewidth]{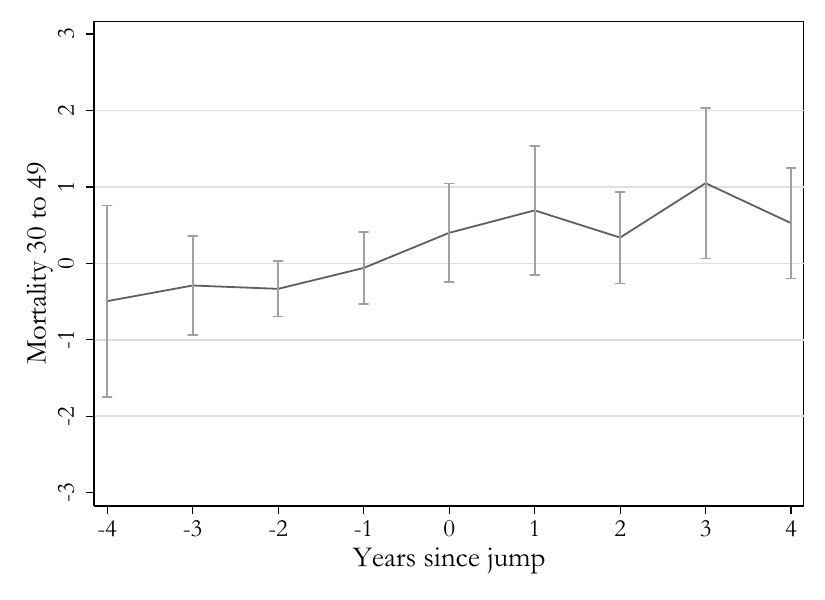}
        \caption{Ages 30 to 49}
    \end{subfigure}
    \begin{subfigure}[b]{0.45\linewidth}
        \centering
        \includegraphics[width=\linewidth]{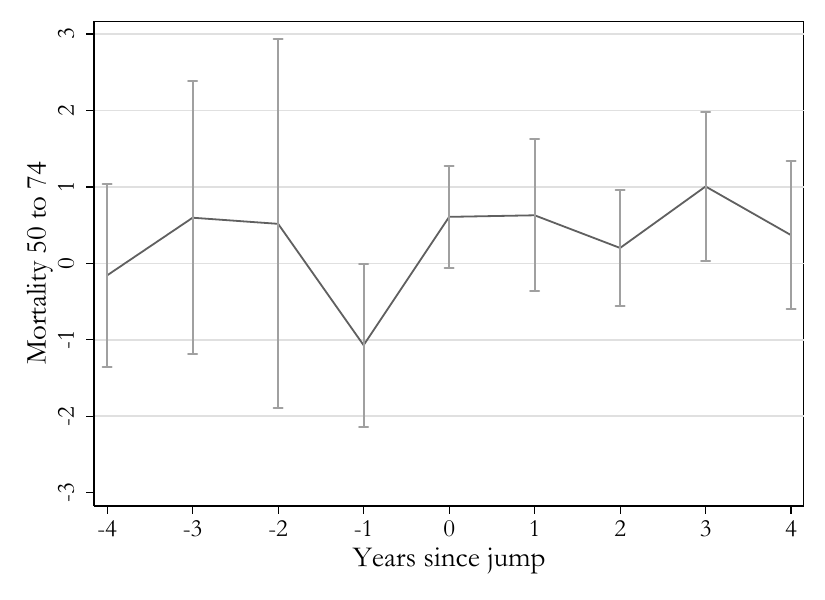}
        \caption{Ages 50 to 74}
    \end{subfigure}
    \begin{subfigure}[b]{0.45\linewidth}
        \centering
        \includegraphics[width=\linewidth]{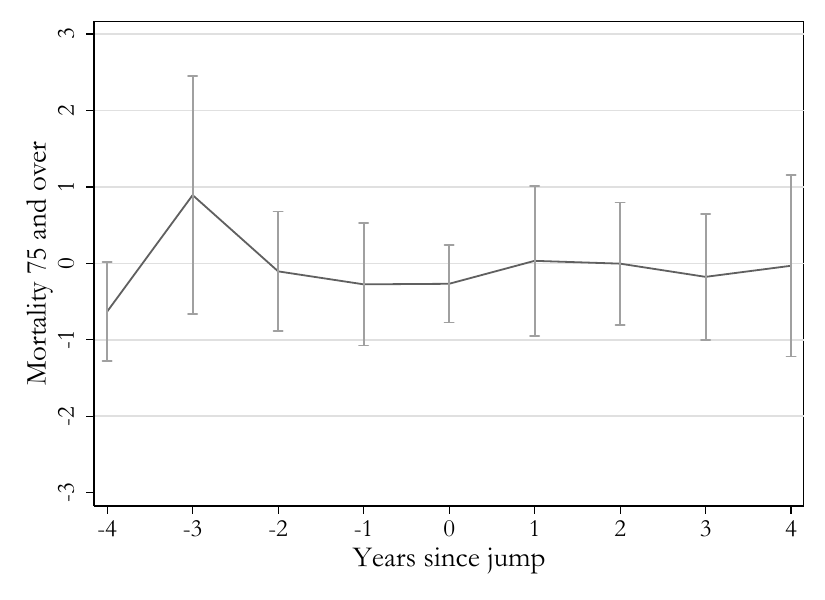}
        \caption{Ages 75-}
    \end{subfigure}
    \caption{Event Study: Effect by Age Group while accounting for Economic Crisis of 1921.}
    \label{fig:mortages1920}
\end{figure}

\end{document}